\documentclass[twocolumn]{aastex631}

\usepackage{amsmath}
\usepackage{natbib}
\usepackage{graphicx}
\usepackage{color}

\usepackage{soul}
\sethlcolor{yellow}
\usepackage[normalem]{ulem}

\def\deg{{$^\circ$}}

\begin{document}

\title{X-ray diagnostics of Cassiopeia A's ``Green Monster'': evidence for dense shocked circumstellar plasma}

\author[0000-0002-4708-4219]{Jacco Vink}
\affiliation{
Anton Pannekoek Institute/GRAPPA, University of Amsterdam, Science Park 904,
1098 XH Amsterdam, The Netherlands
}
\affiliation{
SRON Netherlands Institute for Space Research, Niels Bohrweg 4, 2333 CA Leiden, The Netherlands
}
\author[0000-0001-6965-8642]{Manan Agarwal}
\affiliation{
Anton Pannekoek Institute/GRAPPA, University of Amsterdam, Science Park 904,
1098 XH Amsterdam, The Netherlands
}

\author[0000-0002-6986-6756]{Patrick Slane}
\affiliation{Center for Astrophysics $\vert$ Harvard \& Smithsonian, 60 Garden Street, Cambridge, MA 02138, USA}

\author[0000-0001-9419-6355]{Ilse De Looze}
\affiliation{Sterrenkundig Observatorium, Ghent University, Krijgslaan 281 - S9, B-9000 Gent, Belgium}

\author[0000-0002-0763-3885]{Dan Milisavljevic}
\affiliation{Purdue University, Department of Physics and Astronomy, 525 Northwestern Ave, West Lafayette, IN 47907 }
\affiliation{Integrative Data Science Initiative, Purdue University, West Lafayette, IN 47907, USA}

\author[0000-0002-7507-8115]{Daniel Patnaude}
\affiliation{Center for Astrophysics $\vert$ Harvard \& Smithsonian, 60 Garden Street, Cambridge, MA 02138, USA}

\author[0000-0001-7380-3144]{Tea Temim}
\affiliation{Princeton University, 4 Ivy Ln, Princeton, NJ 08544, USA}

\begin{abstract}
The recent survey of the core-collapse supernova remnant Cassiopeia A (Cas\,A) with the MIRI instrument on board the James Webb Space Telescope (JWST) revealed a large structure in the interior region, referred to as the
``Green Monster". Although its location suggests that it is an ejecta structure, the infrared properties of the ``Green Monster" hint at a circumstellar medium (CSM) origin.
In this companion paper to the JWST Cas A paper, we investigate the filamentary X-ray structures associated with the ``Green Monster" using Chandra X-ray Observatory data. 
We extracted spectra along the ``Green Monster" as well as from shocked CSM regions.
Both the extracted spectra and a principal component analysis show that the ``Green Monster" emission properties are similar to those of the shocked CSM.
The spectra are well-fit by a model consisting of a combination of a non-equilibrium-ionization model and a power-law component, modified by Galactic  absorption.
All the ``Green Monster" spectra show a blueshift corresponding to a radial velocity of around $-2300~{\rm km\ s^{-1}}$, suggesting that the structure is on the near side of Cas A.
The ionization age is around $n_{\rm e}t\approx 1.5\times 10^{11}~{\rm cm^{-3}s}$. This translates into a pre-shock density of $\sim 12~{\rm cm^{-3}}$, higher than previous estimates of the unshocked CSM.
The relatively high $n_{\rm e}t$
and relatively low radial velocity suggest that this structure has a relatively high density compared to other shocked CSM plasma.
This analysis provides yet another piece of evidence that the CSM around  Cas A's progenitor was not that of a smooth steady wind profile.
\end{abstract}


\begin{figure*}[t]
\centerline{\includegraphics[trim=70 0 160 100,clip=true,width=0.96\textwidth]{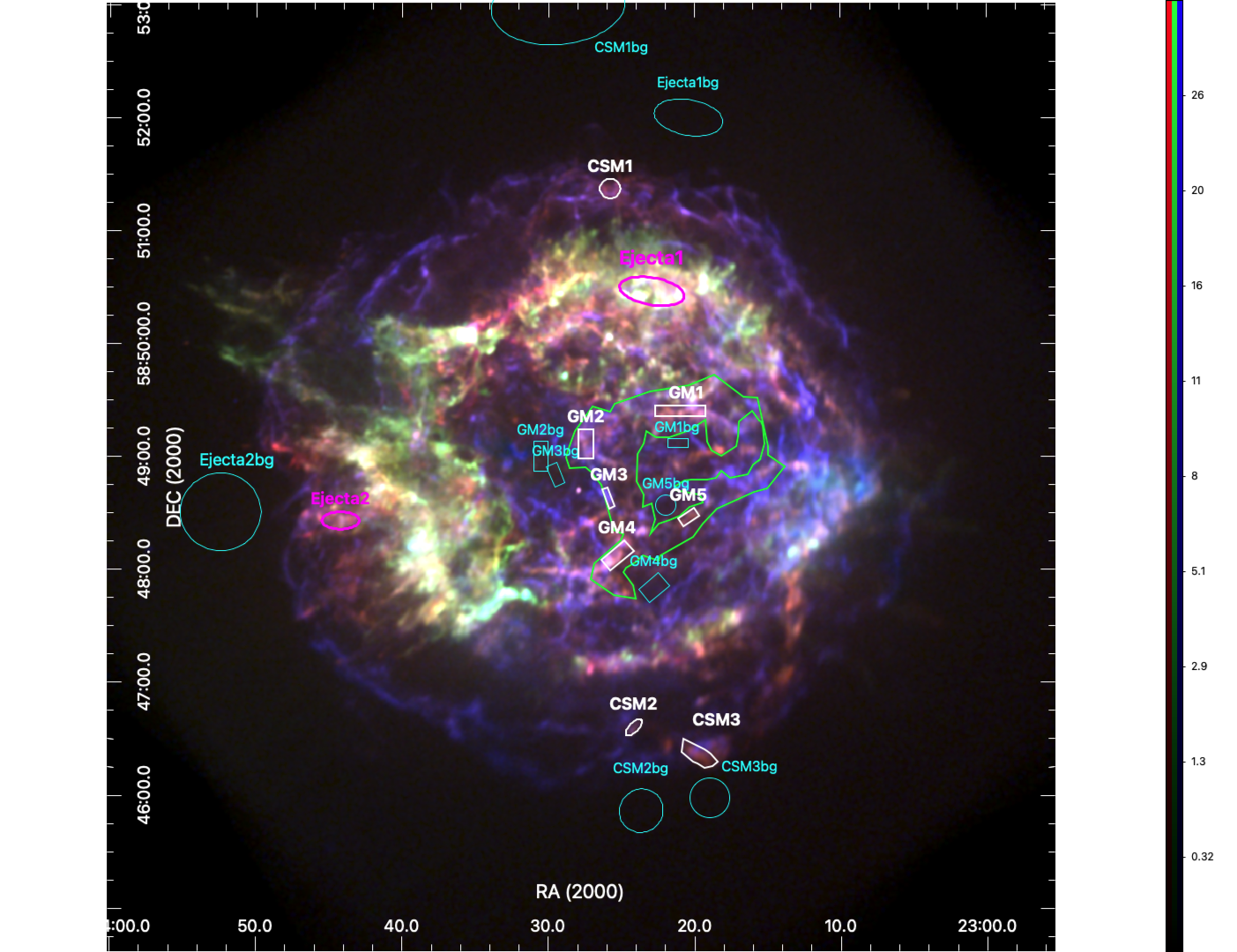}}
\centerline{ \includegraphics[trim=60 0 110 10,clip=true,width=0.32\textwidth]{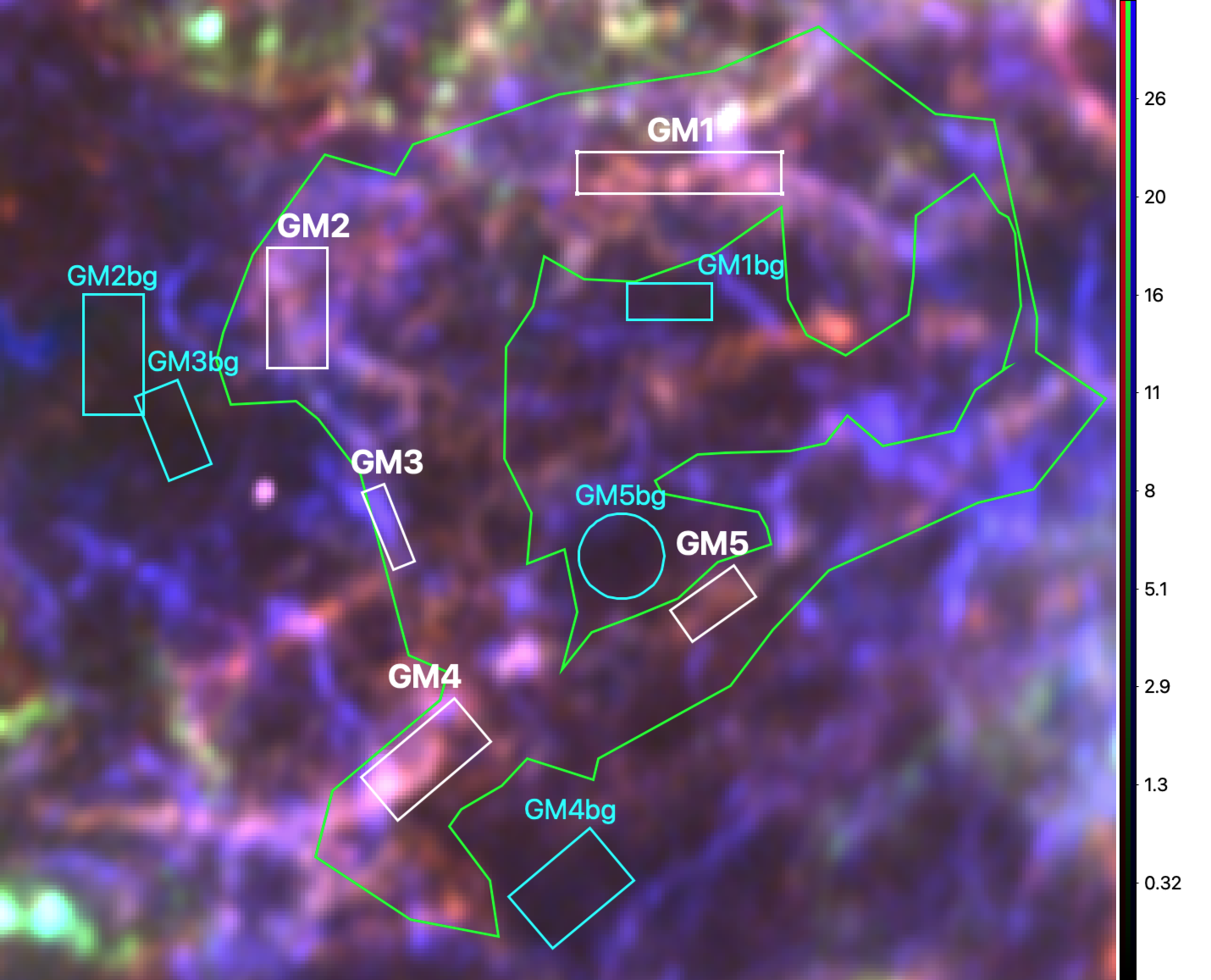}
  \includegraphics[trim=60 0 110 10,clip=true,width=0.32\textwidth]{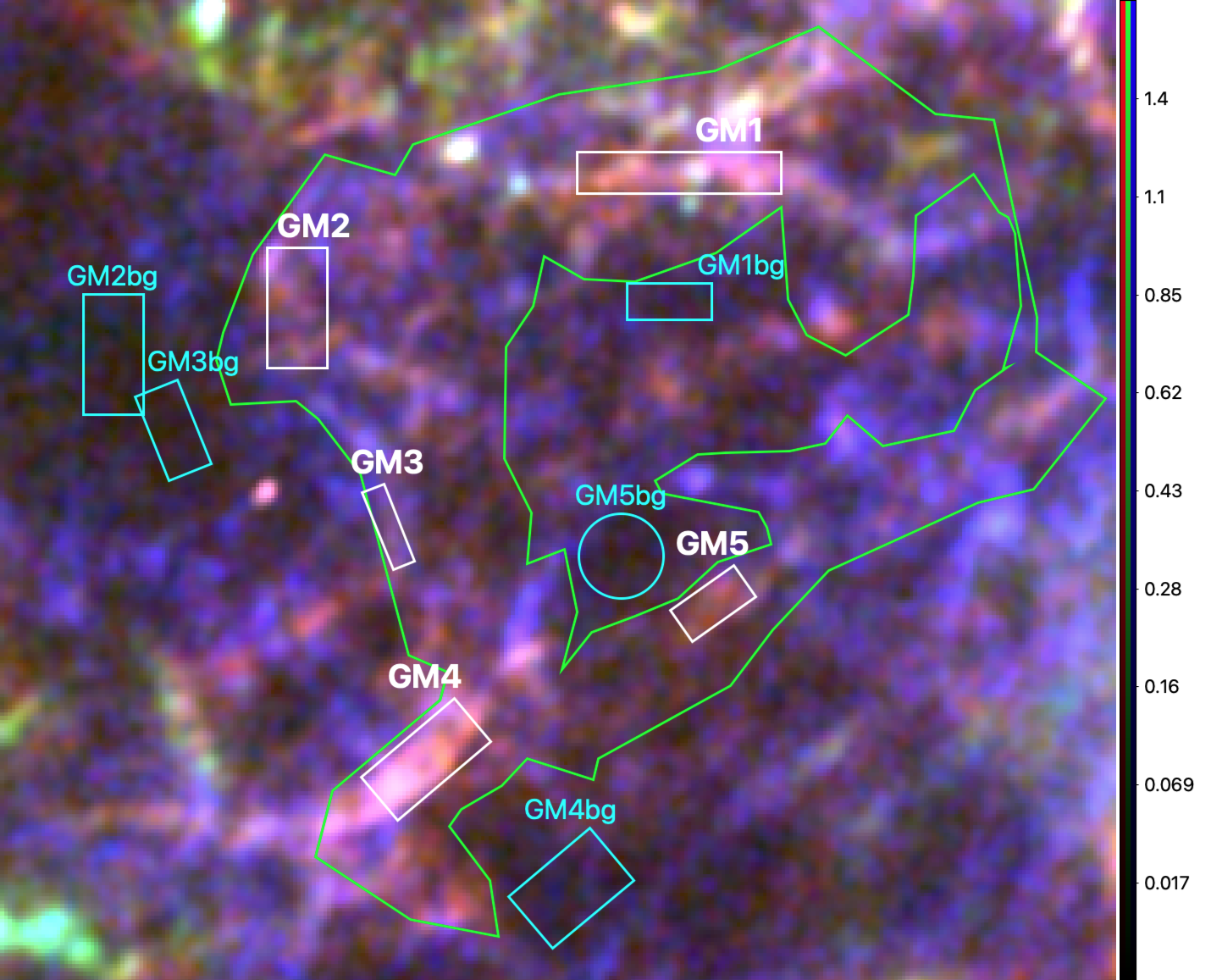}
   \includegraphics[trim=60 0 110 10,clip=true,width=0.32\textwidth]{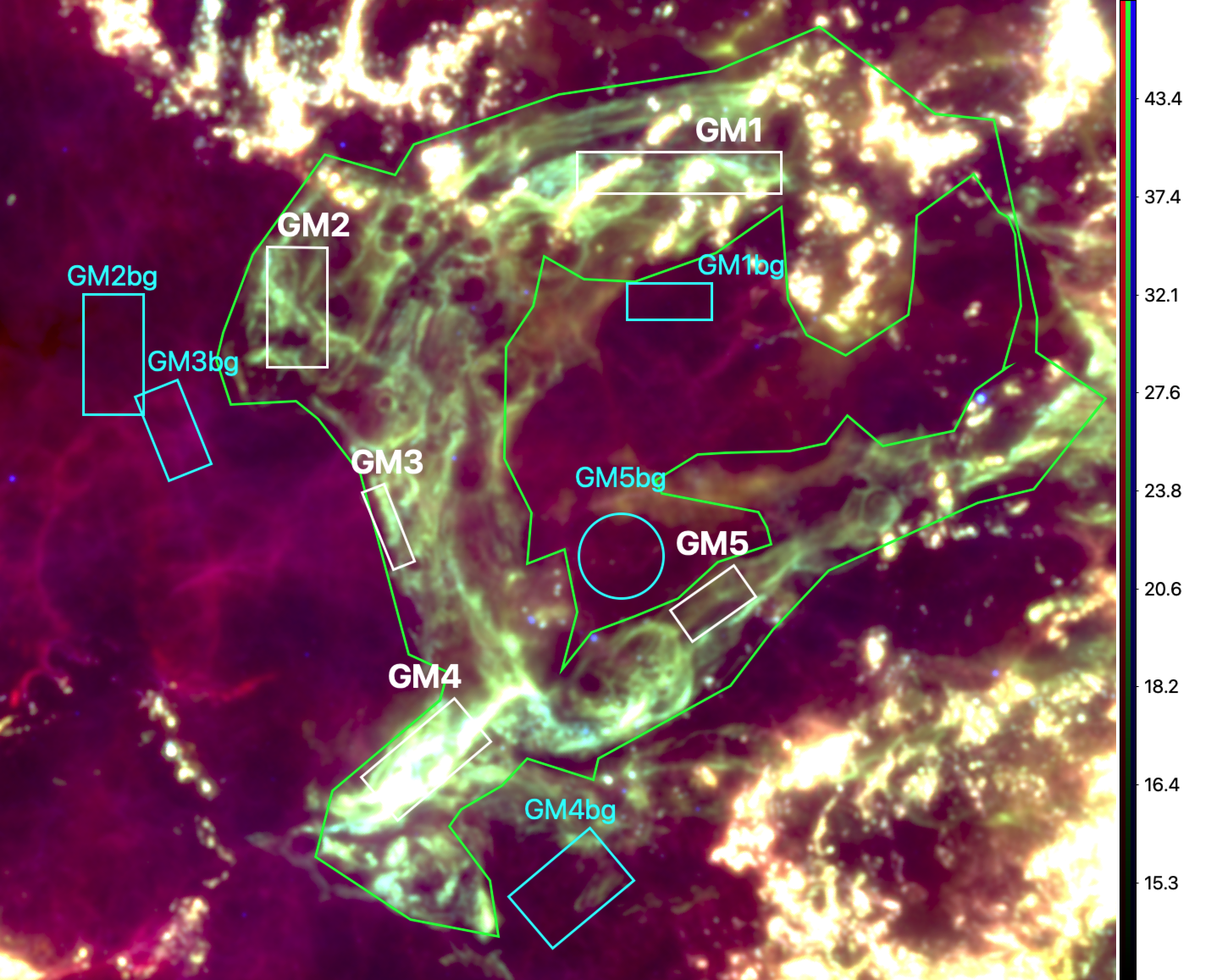}
   }
\caption{
Top panel:
RGB color image using the narrow bands 1.2--1.4~keV (red, showing Fe-L and Mg XI/XII line emission),  1.85-1.98~keV (green, Si XIII), and 4.0--4.5~keV (blue, continuum).
The images are obtained by combining images from several individual deep observations in 2004 (ObsID 4634, 4635, 4636, 4637, 4638, 4723).
The outline of the ``Green Monster" structure is shown with green polygons. Spectral extraction regions along the ``Green Monster" (GM 1--5) and three CSM-related structures (CSM 1--3)
are shown in white. The background regions used are shown in cyan.
Bottom panels: 
The central region of Cas A in X-rays in 2004 (left), 2019 (ObsID19606, center), and IR using JWST/MIRI data (right). The X-ray extraction regions are overlayed as well as a rough outline of the ``Green Monster''.
The RGB channels are for the X-ray images identical to those in the top panel. The JWST map consists of the F2550W (red), F1800W (green), and F1000W (blue) filters \citep[see][]{milisavljevic24}.
All images employ square-root brightness scaling.
\label{fig:casa_regions}
}
\end{figure*}

\section{Introduction}

The infrared imaging mosaics based on the 2022 James Webb Space Telescope  (JWST) observations with the MIRI and NIRCAM camera of the young core-collapse supernova remnant Cassiopeia A (Cas\,A)
 revealed in exquisite detail various spatial infrared (IR)  
components, such as numerous, small ejecta knots emitting thermal dust radiation, a more diffuse circumstellar medium (CSM) component in the outer shell,
and, with NIRCAM, very diffuse emission likely caused by synchrotron radiation \citep{milisavljevic24}.\footnote{\raggedright See also the press release image of April 7, 2023, \url{https://www.nasa.gov/universe/webb-reveals-never-before-seen-details-in-cassiopeia-a}.} 
These basic components were known from previous IR and multiwavelength observations, but 
unexpectedly the JWST images also revealed a giant loop
structure across part of the interior  of Cas A. This IR structure was  popularized under the name the ``Green Monster". 
The location of this loop  interior to the main shell seems to indicate an ejecta origin, but its filamentary structure is morphologically very distinct from the more prominent ejecta components,
consisting of numerous small knots. 

The IR properties  of the ``Green Monster" (GM) are somewhat similar to a large protrusion in the southern part, which is associated with
a bright radio feature \citep[e.g., feature H in][]{braun87}, and with an arc of nitrogen-rich knots, labeled quasi-stationary  flocculi  \citep[QSFs,][]{vandenbergh71,koo20}.  
See  the labeling in Fig.~\ref{fig:casa_legend}.
The spectra of these QSFs show that 
they find their origin in the helium- and nitrogen-enriched  wind of the progenitor star \citep[e.g.,][]{koo23}, and are hence associated with the CSM.

However, its locations and small ``bullet-hole"-like features in the GM filaments prompted us to scrutinize the identification of the GM as a shocked CSM rather than an ejecta structure.
This letter is  a companion paper to the JWST results reported by  \citet{milisavljevic24}---providing a general overview of Cas A observations by JWST--- and the  paper by \citet{delooze24}, which focuses
on  the IR  morphology and spectroscopy of the GM.
\citet{delooze24} report that the IR spectroscopic properties of the GM have similarities with those regions containing QSFs,
and are spectroscopically distinct from other CSM-related dust emission. They attribute this to a high carbon-over-silicate grain abundance ratio for the GM/QSF regions, as compared to other CSM-related dust
components.

Here we show that X-ray filaments associated with the GM do indeed have  X-ray characteristics shared with 
shocked CSM. This analysis is based on  archival X-ray  data gathered by the NASA Chandra X-ray Observatory (Chandra for short).\footnote{Extracted data products as well as software will be made available on \url{ https://doi.org/10.5281/zenodo.10301088}}

\section{X-ray data}

\subsection{X-ray spectra of ``Green Monster" region and the southern CSM}
Chandra observed Cas A numerous times with the ACIS-S CCD detectors in imaging mode. The last such observation concerned a 49ks deep exposure made in 2019 (ObsID 19606).
Much deeper exposures---1~Ms in total---were made in 2004  as part of the Chandra very-large program (VLP) \citep{hwang04}. 

The  X-ray maps using an RGB color representation---see Fig.~\ref{fig:casa_regions}---reveal reddish 
filamentary structures that align well, but not perfectly, with the GM identified in the JWST maps (Fig.~\ref{fig:casa_regions} lower right). \footnote{The JWST data presented in Fig.~\ref{fig:casa_regions} were obtained from the Mikulski Archive for Space Telescopes (MAST) at the Space Telescope Science Institute. The specific observations analyzed can be accessed via \dataset[DOI: 10.17909/szf2-bg42]{https://doi.org/10.17909/szf2-bg42}. 
}
The reddish color in the X-ray maps corresponds with the 
1.2-1.4 keV band, which is characterized by Mg XI/XII, and Fe-L line emission. The reddish color is in stark contrast to the greenish color associated with Si XIII line emission from Si-rich ejecta, 
and the blue/purplish color of X-ray synchrotron radiation. 

The RGB X-ray maps, therefore, already provide an hint that the X-ray emission associated with the GM is not associated with Si-rich ejecta, or synchrotron emission. 
In Fig.~\ref{fig:casa_regions}(top) two other components also appear red:
the iron-rich ejecta in the southeastern region \citep{hughes00a,hwang12}, and a very bright red filament just above the base of the eastern jet, known as ``Baade-Minkowski's filament 1" \citep{baade54}, 
which contains  O/Ne/Mg rich ejecta \citep{vink04a}.

The morphology of these filaments show some evolution between  2004 and 2019 (c.f. Fig.~\ref{fig:casa_regions} bottom left and center). In the north of the GM some stronger Si-lines show up in the 2019 data set, possibly
coinciding with the regions of ejecta knots seen with JWST/MIRI in the northern part of the GM. In addition, the GM overlaps partially with nonthermal X-ray filaments, which seem to  evolve
relatively rapidly in some regions \citep{patnaude11,vink22a}, causing some different spectral characteristics between 2004 and 2019. 
These nonthermal filaments are either associated with the forward shock \citep{gotthelf01a,vink03a}, but projected to the interior, or they are associated with the reverse shock regions \citep{helder08,uchiyama08}, requiring relative speeds between the reverse shock and the unshocked ejecta of $\gtrsim 3000~{\rm km\ s^{-1}}$, a requirement that seems to met in parts of the western half of Cas A \citep{vink22a}.
Since the reddish-colored filaments associated with the GM itself are relatively stable between 2004 and 2019, we opted for an investigation of the
deeper 2004 VLP observation, by combining all nine ObsIDs comprising the 1Ms observation, contained in ~\dataset[DOI: 10.25574/cdc.209]{https://doi.org/10.25574/cdc.209}.

We extracted five spectra along the GM filaments, as well as a region associated at the tip of the arc of  QSFs (CSM3), a fainter region in the south, outside the ejecta shell  (CSM2), and also associated
with QSFs. For comparison we also extracted a relatively bright CSM region in the north (CSM1), with no QSF counterpart.
Background spectra were obtained from faint nearby regions, assuming that the extracted spectra of filaments contain also more diffuse components plus other background components.
We note here that even background spectra from extraction regions outside Cas A still seem to be dominated by dust scattered X-ray emission from Cas A  itself, 
but also by ``out of time" events (i.e. photons detected during readout time, causing these photons to be assigned wrong sky coordinates).
All extraction regions are displayed in Fig.~\ref{fig:casa_regions}.
All the spectra were extracted individually from each ObsID and then summed using the {\tt combine\_spectra} script in the Chandra software package ciao-4.15.
Subsequently
the spectra were optimally binned following the procedure described in \cite{Kaastra16} through the FTool \citep{ftools} task {\tt ftgrouppha}.

For fitting spectra, we experimented with both a model using the standard  non-equilibrium ionization (NEI) model {\tt vnei} and the  {\tt vpshock} model  \citep{borkowski01}. The {\tt vnei} 
assumes a single ionization age $\tau$ (or $n_{\rm e}t$), whereas {\tt vpshock} assumes a superposition of ionization ages assuming a plane parallel shock geometry. 
Since the  {\tt vpshock} is more realistic, we report here only the {\tt vpshock} model fits. The {\tt vpshock} model specifies a lower and an upper bound to the ionization age,  $\tau_{\rm max}$. 
The lower bound was set to zero, and the upper bound
was left as a free parameter.
Apart from the NEI component the model includes Galactic absorption \citep[{\tt tbabs},][]{wilms00}, and a power-law component to account for the X-ray synchrotron radiation.
The abundances were not well constrained for all spectra, in particular for those spectra that showed evidence for a prominent synchrotron component  (GM3). For that reason we 
made simultaneous fits to all GM spectra, coupling the abundances across all the GM spectra, but determining the other relevant parameters for each individual spectrum. We used the same procedure for the three
CSM spectra.  Apart from the electron temperature $kT_{\rm e}$, $\tau_{\rm max}$, the normalizations  and absorption columns $N_{\rm H}$, we also allowed for line Doppler shifts.
The abundance of Ni was coupled to Fe, and the abundance of Ca was coupled to Ar. The H, He, C, and N abundances were fixed to solar abundances, and the rest were free to vary between 0.1 and 5 with log-uniform priors. All abundances were measured in units of the \citet{lodders09} proto-solar abundance values.

For the optimization of the fitted model, we used the Bayesian scheme employed in \citet{ellien23}, based on the Python package Bayesian X-ray Analysis \citep[BXA v4.1.1;][]{buchner14} which links the nested sampling algorithm, {\tt Ultranest} \citep{buchner21}, with {\tt PyXspec}, a Python implementation of the classic X-ray analysis package {\tt xspec} \citep[v12.13.1;][]{arnaud96}. Nested sampling is a Monte Carlo technique that ensures extensive coverage of the parameter space \citep{ashton22, buchner23}. In this work, we used the Reactive Nested Sampler in {\tt Ultranest} with 800 live points and a log-evidence accuracy of 0.5. A simultaneous fit to all combined GM, and combined CSM spectra was performed using BXA with 42 and 28 free parameters, respectively.

\begin{figure*}
\includegraphics[trim=0 0 0 0,angle=0,clip=true,width=0.5\textwidth]{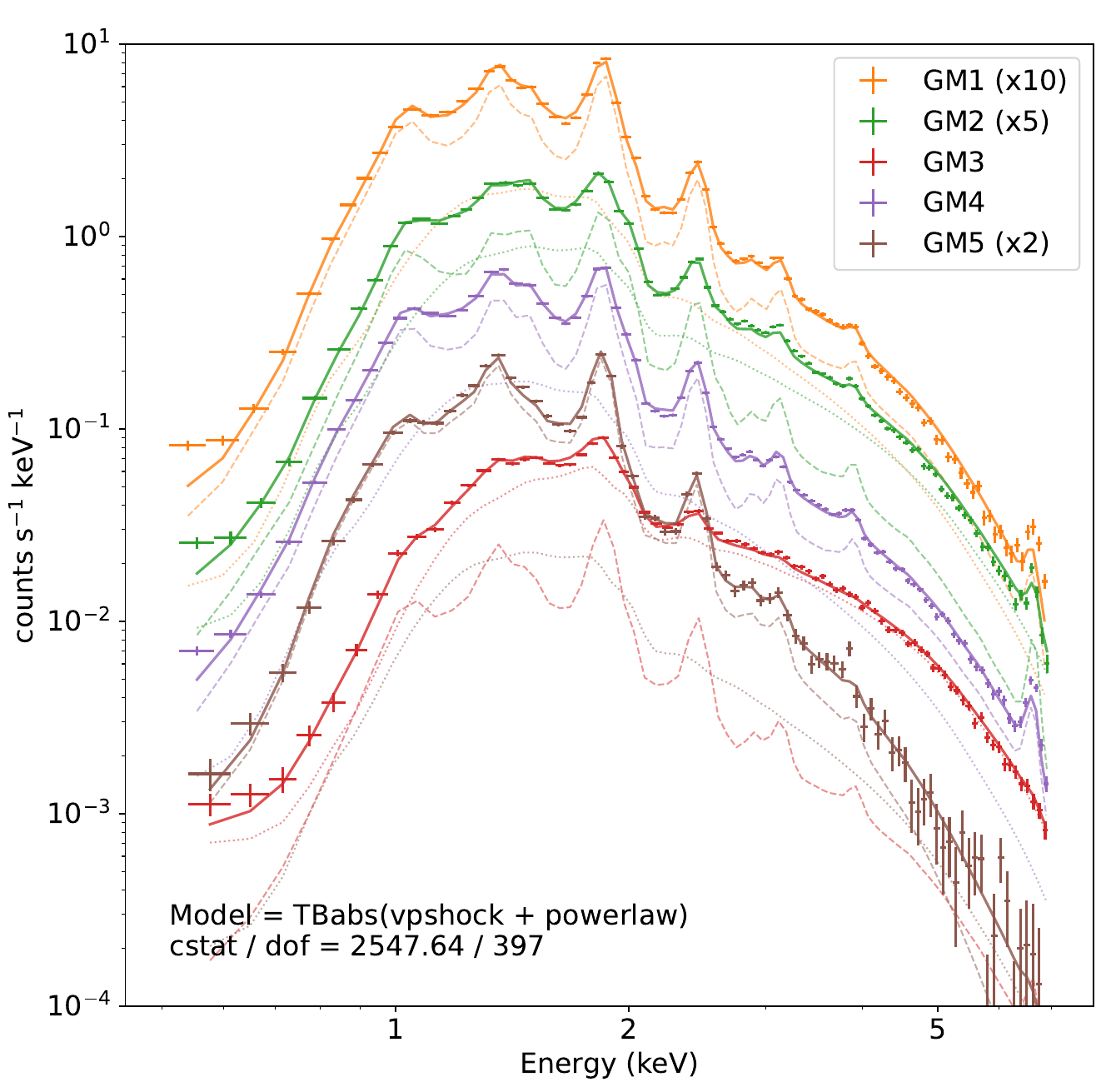}
\includegraphics[trim=0 0 0 0,angle=0,clip=true,width=0.5\textwidth]{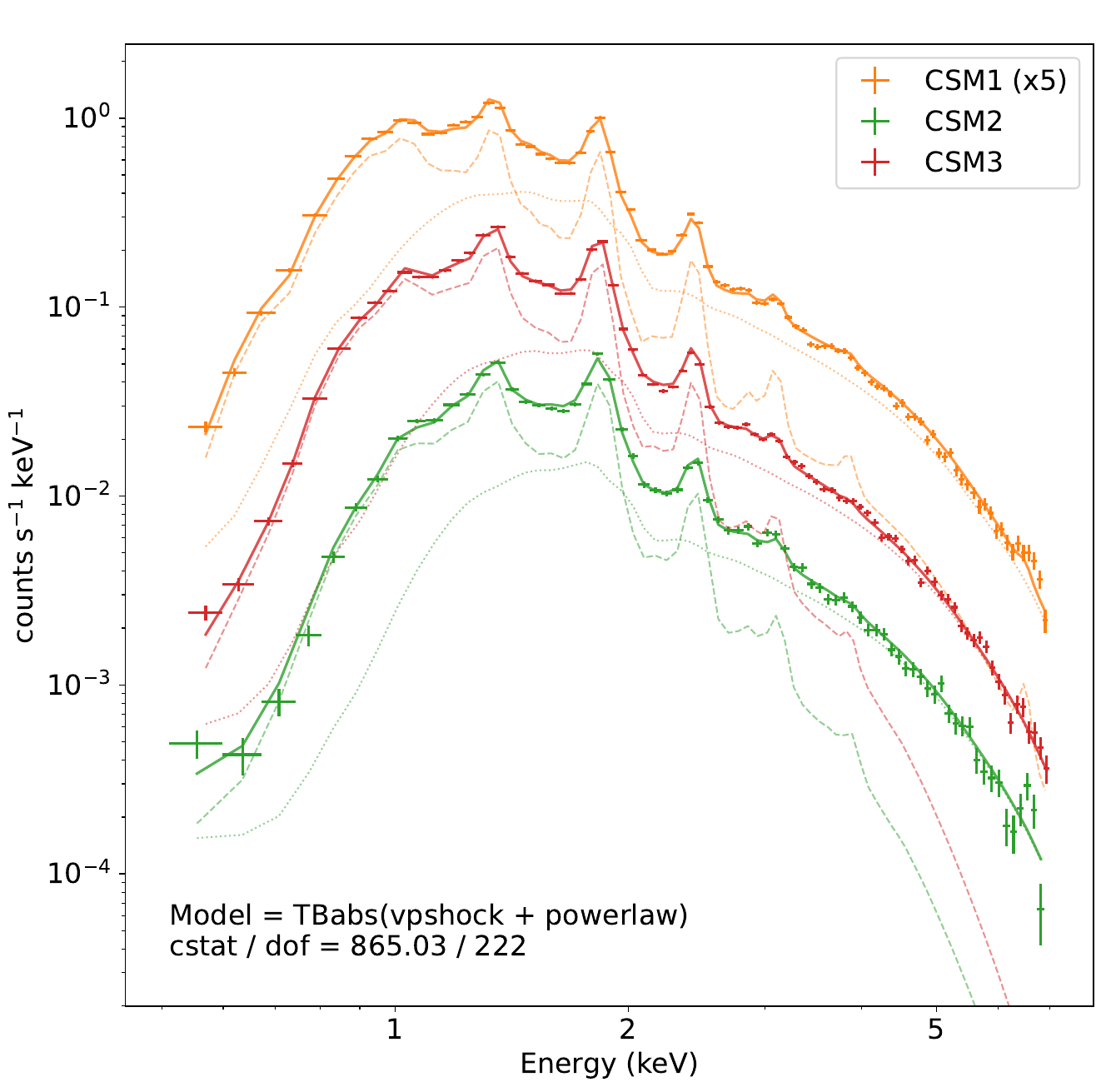}
\caption{Left:
X-ray spectra from the ``Green Monster" region. 
Right: Spectra of the three CSM regions. 
See Fig.~\ref{fig:casa_regions} and Table~\ref{tab:gm_spectra} for the identification and spectral fit parameters. The dashed and dotted lines show the model contributions
of the {\tt vpshock} and powerlaw components, respectively.
\label{fig:spectra}
}
\end{figure*}

All fitted models provide reasonable fit to the spectra,  given that these are line-rich spectra with excellent statistics. The latter implies that systematic uncertainties dominate.
The spectra and best-fit models are shown in Fig.~\ref{fig:spectra}.
The priors we used and the best-fit model parameters are listed in Table.~\ref{tab:gm_spectra}. The stated parameter values represent the maximum likelihood value and 1 $\sigma$ credible interval given the corresponding posterior distribution. See \citet{eadie23} for definitions of maximum likelihood and credibility interval in Bayesian statistics. We note that some of the estimated credibility intervals are rather small, and should be used with caution as the systematic errors likely dominate.

\begin{table*}
\caption{Best-fit parameters for GM and CSM spectra.
\label{tab:gm_spectra}}
\scriptsize
\hspace*{-2cm}\begin{tabular}{c|c|ccccc|ccc}\hline\hline\noalign{\smallskip}
				&	Priors	&		GM 1						&		GM 2						&		GM 3						&		GM 4						&		GM 5						&		CSM 1						&		CSM 2						&		CSM 3						\\	\noalign{\smallskip}\hline\noalign{\smallskip}
$kT_{\rm e}$ (keV)				&	0.1--5\tablenotemark{\small c}	&	$	1.92	^{+	0.03	}_{-	0.03	}$	&	$	2.19	^{+	0.06	}_{-	0.06	}$	&	$	2.33	^{+	0.33	}_{-	0.33	}$	&	$	3.01	^{+	0.06	}_{-	0.06	}$	&	$	0.88	^{+	0.02	}_{-	0.02	}$	&	$	1.76	^{+	0.12	}_{-	0.12	}$	&	$	0.79	^{+	0.06	}_{-	0.06	}$	&	$	0.78	^{+	0.03	}_{-	0.03	}$	\\	
O				&	0.1--5	&	\multicolumn{5}{c|}{ $	4.97	^{+	0.05	}_{-	0.05	}$ }					&	\multicolumn{3}{c}{ $	4.56	^{+	0.36	}_{-	0.33	}$ }			\\	
Ne				&	0.1--5	&	\multicolumn{5}{c|}{ $	2.67	^{+	0.07	}_{-	0.07	}$ }					&	\multicolumn{3}{c}{ $	2.60	^{+	0.24	}_{-	0.22	}$ }			\\	
Mg				&	0.1--5	&	\multicolumn{5}{c|}{ $	1.92	^{+	0.04	}_{-	0.04	}$ }					&	\multicolumn{3}{c}{ $	1.85	^{+	0.16	}_{-	0.15	}$ }			\\	
Si				&	0.1--5	&	\multicolumn{5}{c|}{ $	1.98	^{+	0.05	}_{-	0.05	}$ }					&	\multicolumn{3}{c}{ $	1.72	^{+	0.17	}_{-	0.15	}$ }			\\	
S				&	0.1--5	&	\multicolumn{5}{c|}{ $	1.92	^{+	0.05	}_{-	0.05	}$ }					&	\multicolumn{3}{c}{ $	2.50	^{+	0.29	}_{-	0.26	}$ }			\\	
Ar				&	0.1--5	&	\multicolumn{5}{c|}{ $	1.73	^{+	0.07	}_{-	0.07	}$ }					&	\multicolumn{3}{c}{ $	3.86	^{+	0.66	}_{-	0.56	}$ }			\\	
Fe				&	0.1--5	&	\multicolumn{5}{c|}{ $	1.41	^{+	0.04	}_{-	0.04	}$ }					&	\multicolumn{3}{c}{ $	1.49	^{+	0.11	}_{-	0.10	}$ }			\\	
$\tau_{\rm max} (10^{11}{\rm cm^{-3}s})$				&	0.05--8	&	$	1.17	^{+	0.03	}_{-	0.03	}$	&	$	2.24	^{+	0.10	}_{-	0.10	}$	&	$	0.98	^{+	0.14	}_{-	0.12	}$	&	$	0.96	^{+	0.02	}_{-	0.02	}$	&	$	3.56	^{+	0.24	}_{-	0.22	}$	&	$	0.62	^{+	0.04	}_{-	0.04	}$	&	$	2.66	^{+	0.56	}_{-	0.46	}$	&	$	2.32	^{+	0.22	}_{-	0.20	}$	\\	
$v_{\rm rad} ({\rm km\,s^{-1}})$				&	-3000--3000\tablenotemark{\small c}	&	$	-2,400	^{+	3	}_{-	3	}$	&	$	-1,240	^{+	56	}_{-	56	}$	&	$	-2,886	^{+	163	}_{-	163	}$	&	$	-1,878	^{+	11	}_{-	11	}$	&	$	-2,314	^{+	26	}_{-	26	}$	&	$	-61	^{+	48	}_{-	48	}$	&	$	-1,453	^{+	212	}_{-	212	}$	&	$	24	^{+	84	}_{-	84	}$	\\	
norm \tablenotemark{\small a}				&	0.1--100	&	$	3.76	^{+	0.13	}_{-	0.12	}$	&	$	1.63	^{+	0.05	}_{-	0.05	}$	&	$	0.19	^{+	0.02	}_{-	0.01	}$	&	$	2.60	^{+	0.08	}_{-	0.08	}$	&	$	1.19	^{+	0.04	}_{-	0.04	}$	&	$	0.56	^{+	0.07	}_{-	0.06	}$	&	$	0.61	^{+	0.08	}_{-	0.07	}$	&	$	2.14	^{+	0.20	}_{-	0.18	}$	\\	
$\Gamma$				&	2.5--3.7\tablenotemark{\small c}	&	$	3.69	^{+	0.01	}_{-	0.01	}$	&	$	2.95	^{+	0.01	}_{-	0.01	}$	&	$	2.56	^{+	0.02	}_{-	0.02	}$	&	$	3.70	^{+	0.01	}_{-	0.01	}$	&	$	3.69	^{+	0.03	}_{-	0.03	}$	&	$	2.89	^{+	0.03	}_{-	0.03	}$	&	$	2.98	^{+	0.07	}_{-	0.07	}$	&	$	2.93	^{+	0.05	}_{-	0.05	}$	\\	
PL norm \tablenotemark{\small b}				&	$10^{-7}-10^{2}$	&	$	5.26	^{+	0.13	}_{-	0.13	}$	&	$	3.45	^{+	0.06	}_{-	0.06	}$	&	$	1.27	^{+	0.03	}_{-	0.03	}$	&	$	4.81	^{+	0.10	}_{-	0.10	}$	&	$	0.38	^{+	0.03	}_{-	0.03	}$	&	$	1.15	^{+	0.04	}_{-	0.04	}$	&	$	0.39	^{+	0.04	}_{-	0.04	}$	&	$	1.24	^{+	0.11	}_{-	0.10	}$	\\	
$N_{\rm H} ({\rm cm^{-2}})$				&	0.1--4\tablenotemark{\small c}	&	$	1.86	^{+	0.01	}_{-	0.01	}$	&	$	1.72	^{+	0.01	}_{-	0.01	}$	&	$	2.15	^{+	0.02	}_{-	0.02	}$	&	$	1.77	^{+	0.01	}_{-	0.01	}$	&	$	2.04	^{+	0.01	}_{-	0.01	}$	&	$	1.35	^{+	0.02	}_{-	0.02	}$	&	$	2.20	^{+	0.03	}_{-	0.03	}$	&	$	1.82	^{+	0.02	}_{-	0.02	}$	\\	
C-stat/bins				&		&		896.96	/	91				&		503.43	/	91				&		190.94	/	83				&		710.08	/	92				&		246.23	/	82				&		402.48	/	91				&		139.62	/	74				&		322.92	/	85				\\	\noalign{\smallskip}\hline\noalign{\smallskip}

\end{tabular}
\tablenotetext{a}{\footnotesize Defined as norm$=$EM$=10^{-14}/(4\pi d^2)\int n_{\rm e}n_{\rm H}dV$.}
\tablenotetext{b}{\footnotesize units: $10^{-3}{\rm ph\,cm^{-2}\,keV^{-1}}$ @ 1 keV}
\tablenotetext{c}{\footnotesize For these particular parameters uniform priors in linear space were assumed.
For all other parameters, log-uniform priors were assumed.}
\end{table*}

\begin{figure}
 \includegraphics[width=\columnwidth]{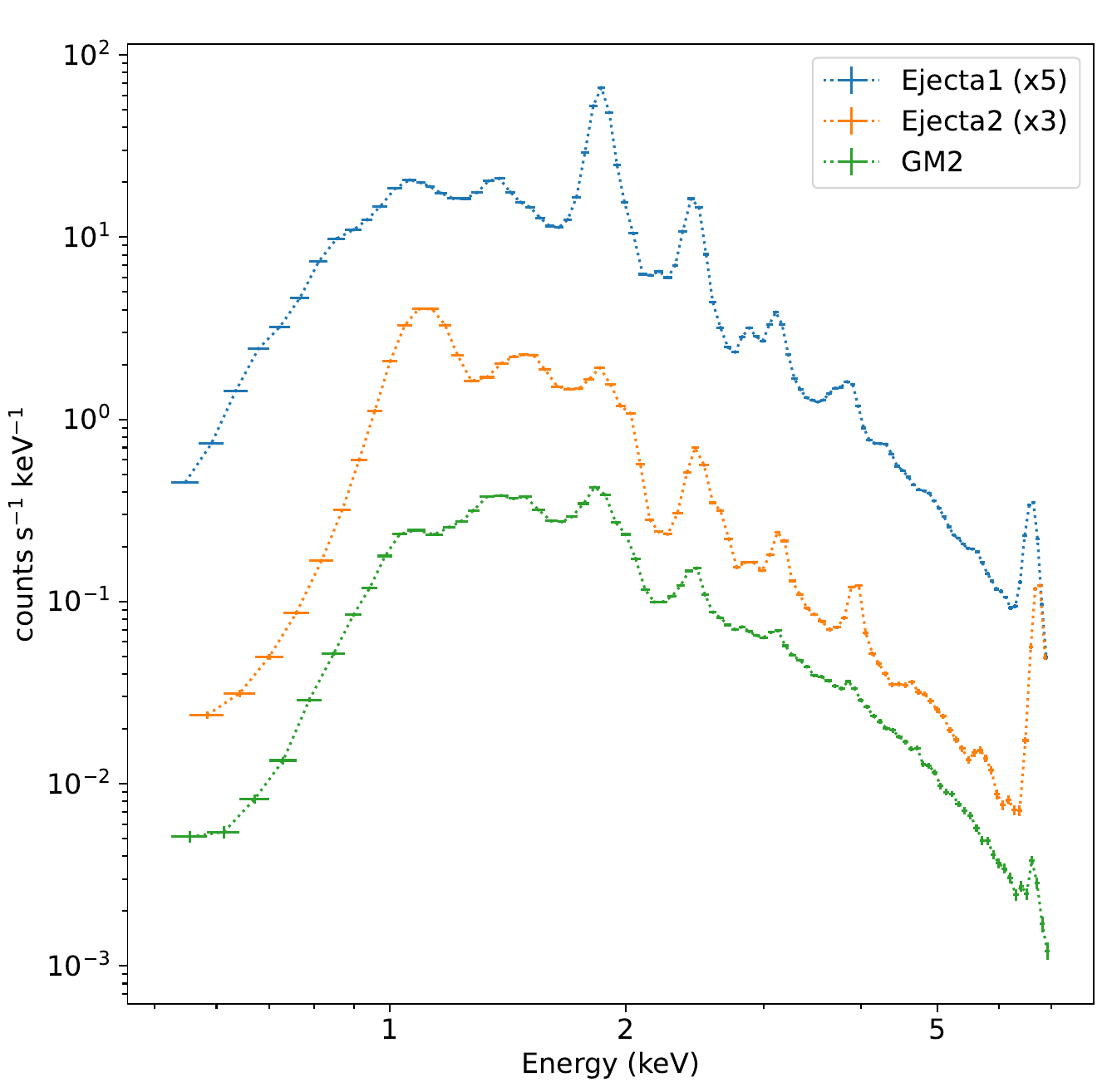}
    \caption{
   A comparison of two spectra from ejecta-dominated regions with one of the GM spectra (GM2).  The extraction regions are indicated in Fig.~\ref{fig:casa_regions}. Ejecta1 corresponds to Si-rich
   ejecta and Ejecta2 corresponds to Fe-rich ejecta. All spectra are background subtracted and optimally binned.
   A dotted line is drawn between the spectral data points for clarity.
    \label{fig:ejecta}}
\end{figure}

\subsection{Principal component analysis}
The GM spectra are characterized by less-prominent Si-K and S-K lines as compared to ejecta spectra, which have either bright Si-K, S-K lines,
or otherwise prominent Fe-L and Fe-K line complexes \citep[e.g.][]{hughes00a,hwang12}. 
We illustrate this in Fig.~\ref{fig:ejecta}, where we show spectra from two ejecta-dominated regions in comparison with the GM2 spectrum.
However, a low equivalent width of the Si-K and S-K lines can also be caused by a dominant X-ray synchrotron composition. So it is difficult
from spectral fitting alone to establish whether the GM spectra are originating from metal-rich ejecta plasma, with  a strong nonthermal component,
or originate from CSM. 

In order to better characterize the X-ray spectral content of Cas A, we resorted to a principal component analysis (PCA).
PCA is a well-known statistical method in which a complex data set, represented as a set of data vectors, is restructured into a set of
orthogonal eigen vectors, ordered according to their relative importance, i.e., the amount of variance they cover.
PCA has been used before to characterize X-ray spectra of young SNRs, see \citet{warren04,warren06}.

The PCA of the data consists of representing the spatial-plus-energy data cube as a two dimensional matrix, with the
rows corresponding to all the spatial pixels and the columns corresponding to discrete energy bands. For this PCA,
we combined the ObsIDs 4634, 4635, 4636, 4637, 4638, 4723, creating 28 combined images (Table~\ref{tab:pca}) with a pixel resolution of 0.98\arcsec. 
The images were  rebinned to a pixel size of  
1.97\arcsec\ (4 by 4 Chandra pixels).
The columns were centered (mean per band subtracted) and divided by the square root of the pixel variance for each band.
The PCA output consists of the eigenvectors (principal components, PCs), 
the singular values, and the PC scores. The latter are images indicating for each pixel the amount by which it is represented by a PC. 

\begin{figure*}
\centerline{
\includegraphics[trim=300 240 350 300,clip=true,width=0.24\textwidth]{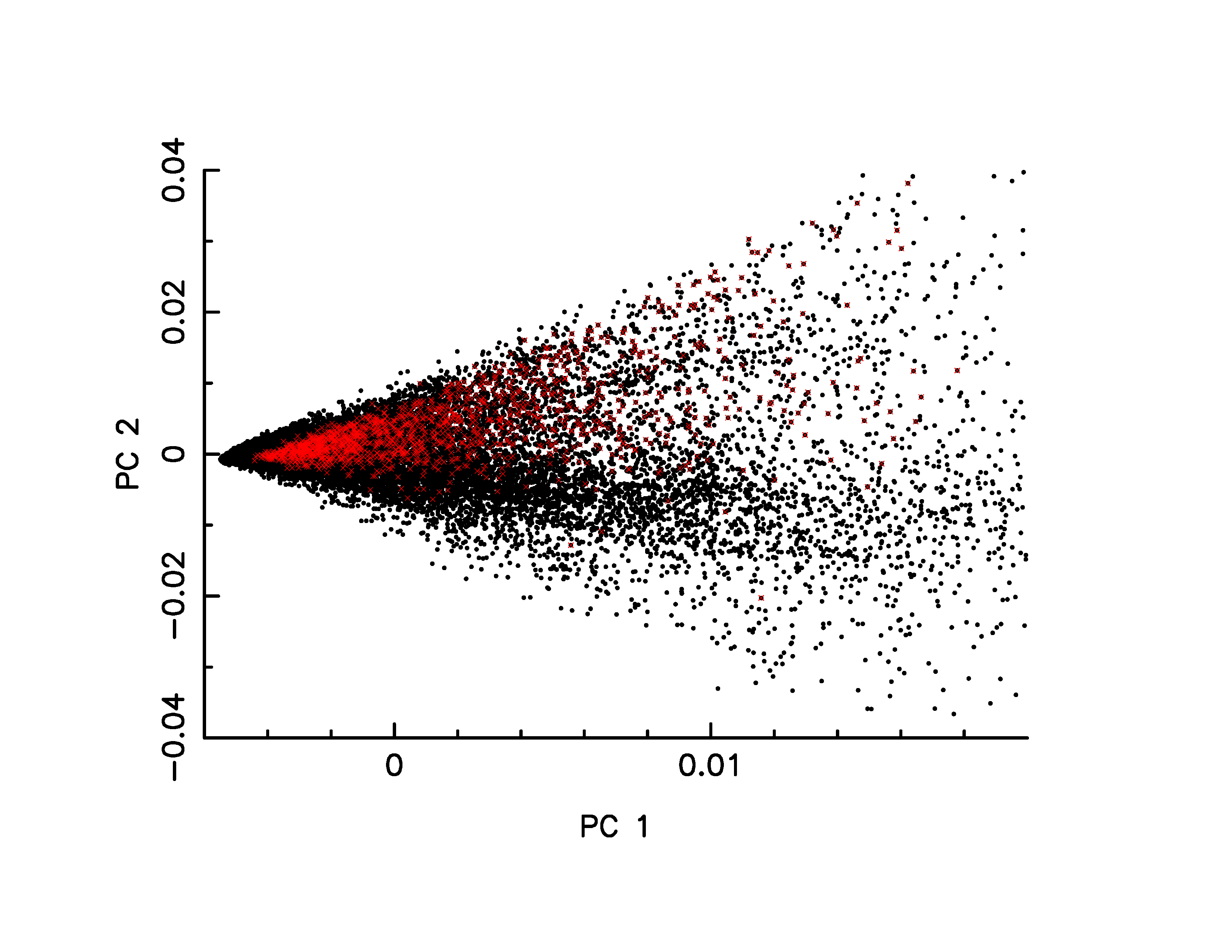}
\includegraphics[trim=300  240 350 300,clip=true,width=0.24\textwidth]{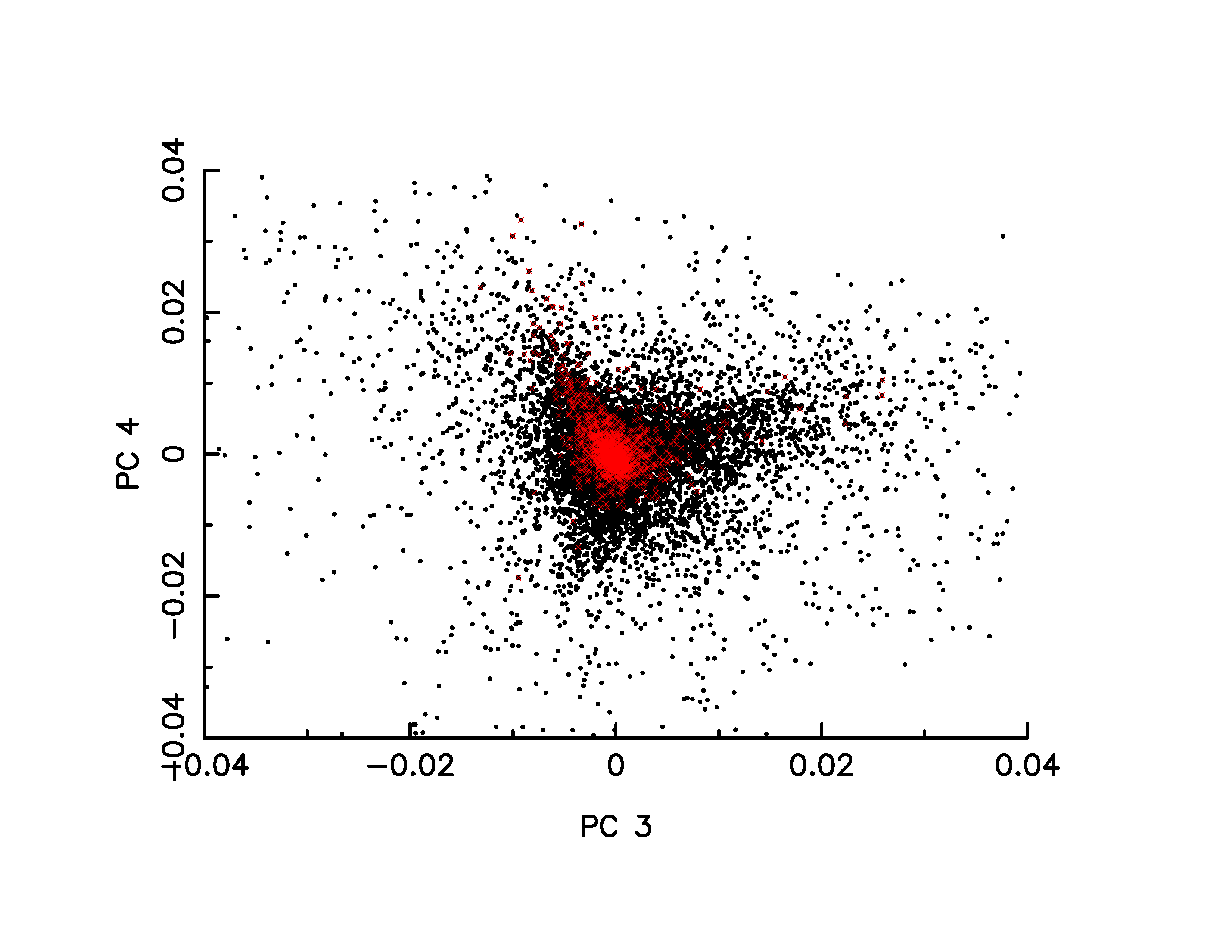}
\includegraphics[trim=300  240 350 300,clip=true,width=0.24\textwidth]{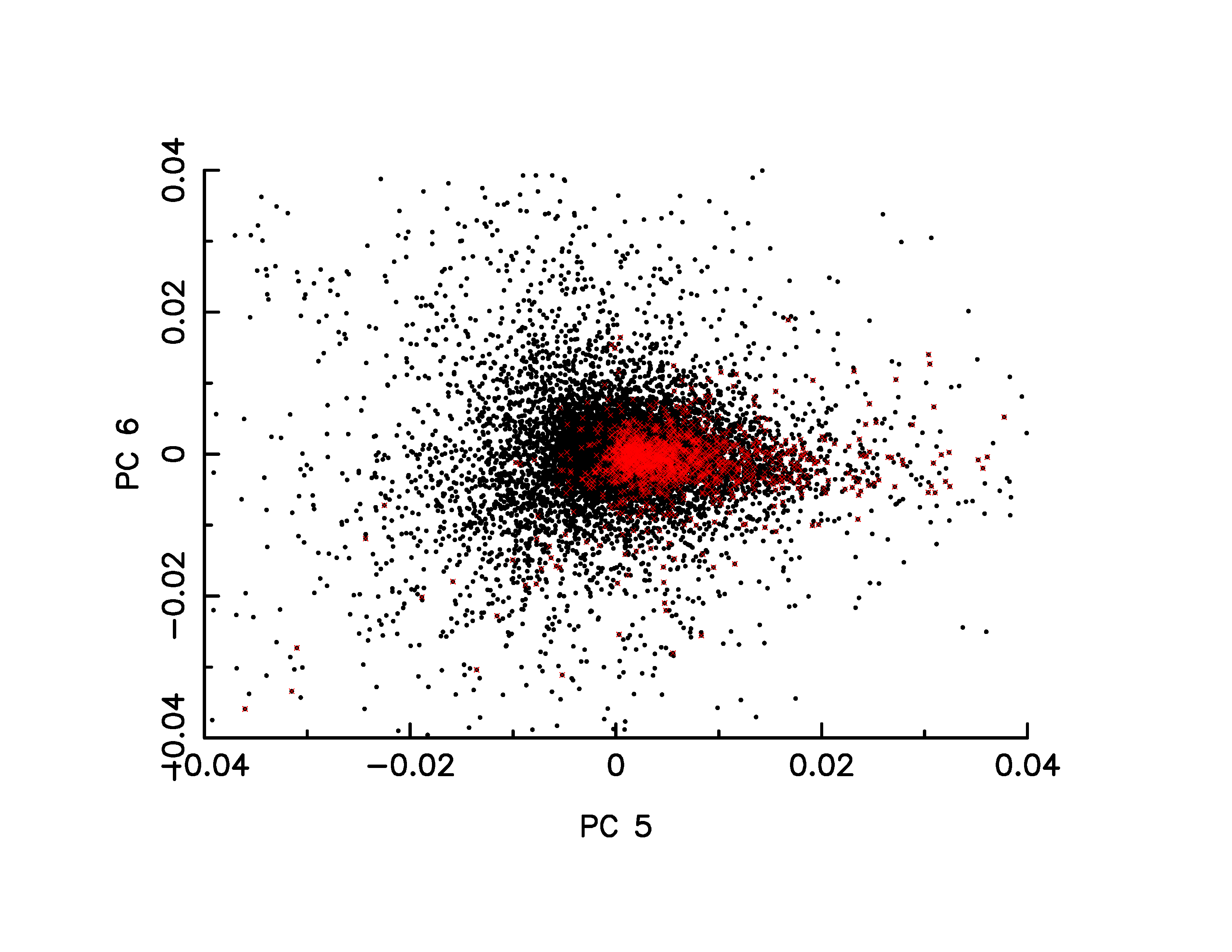}
\includegraphics[trim=300  240 350 300,clip=true,width=0.24\textwidth]{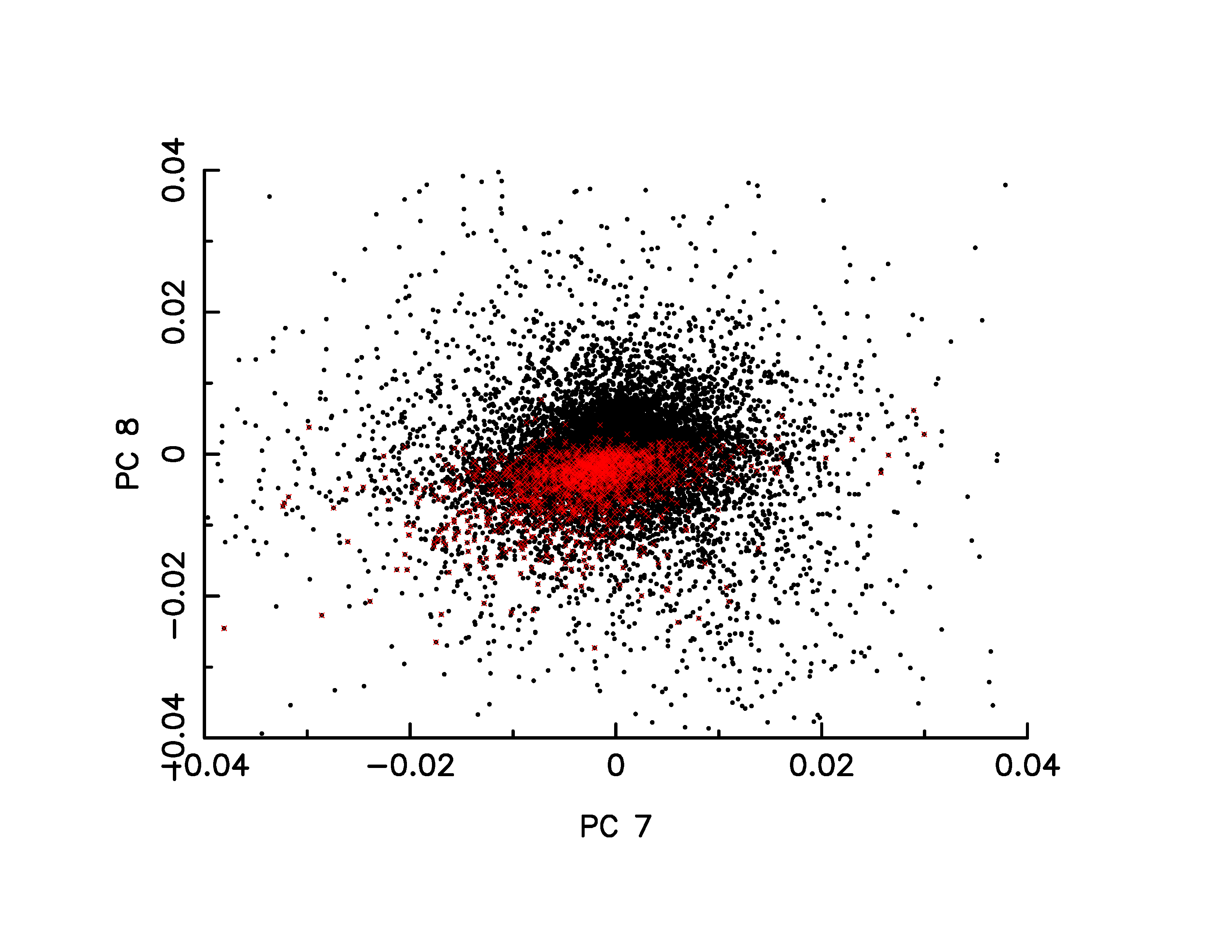}
}
\centerline{
\includegraphics[trim=50 0 150 50,clip=true,width=0.5\textwidth]{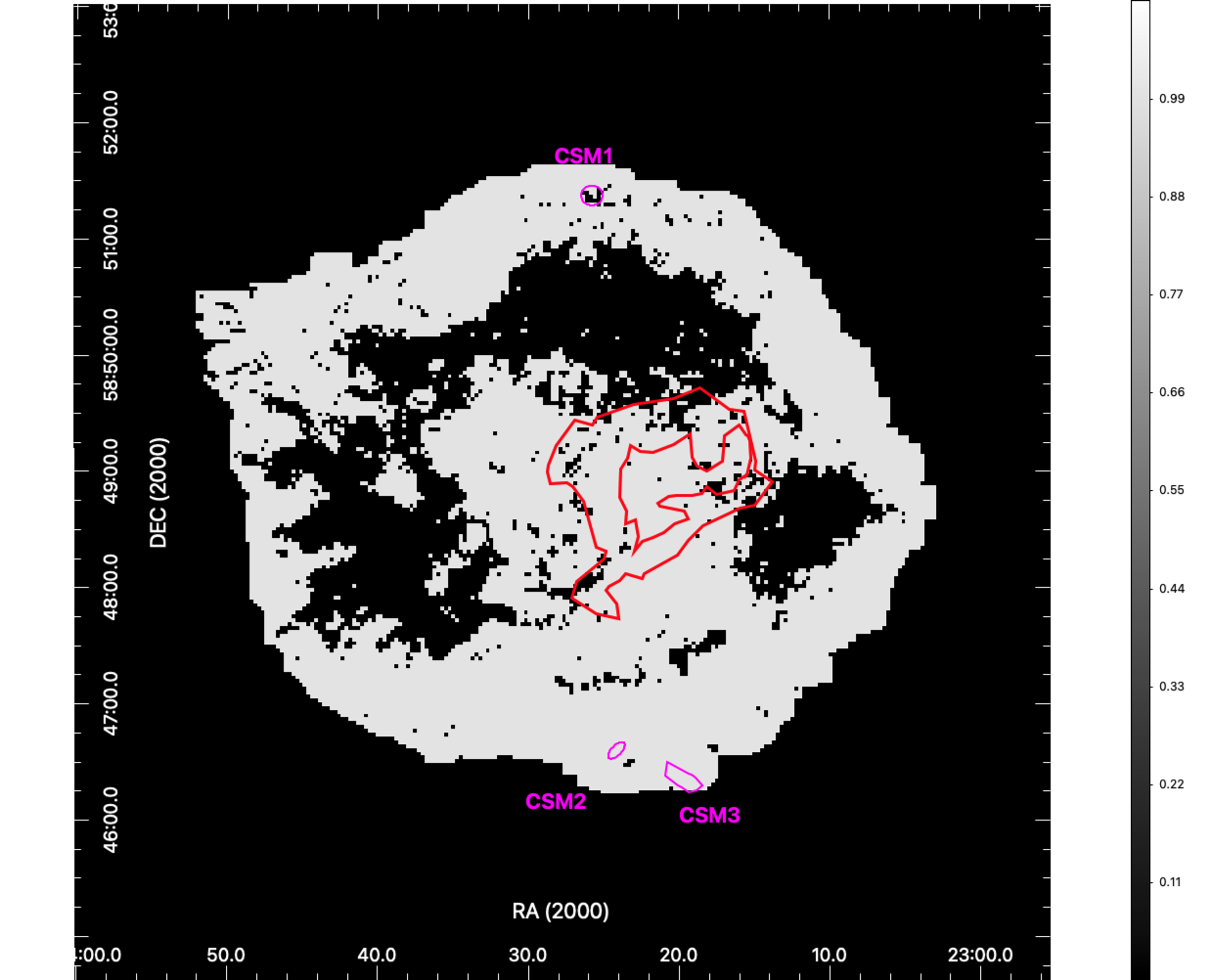}
\includegraphics[trim=50 0 150 50,clip=true,width=0.5\textwidth]{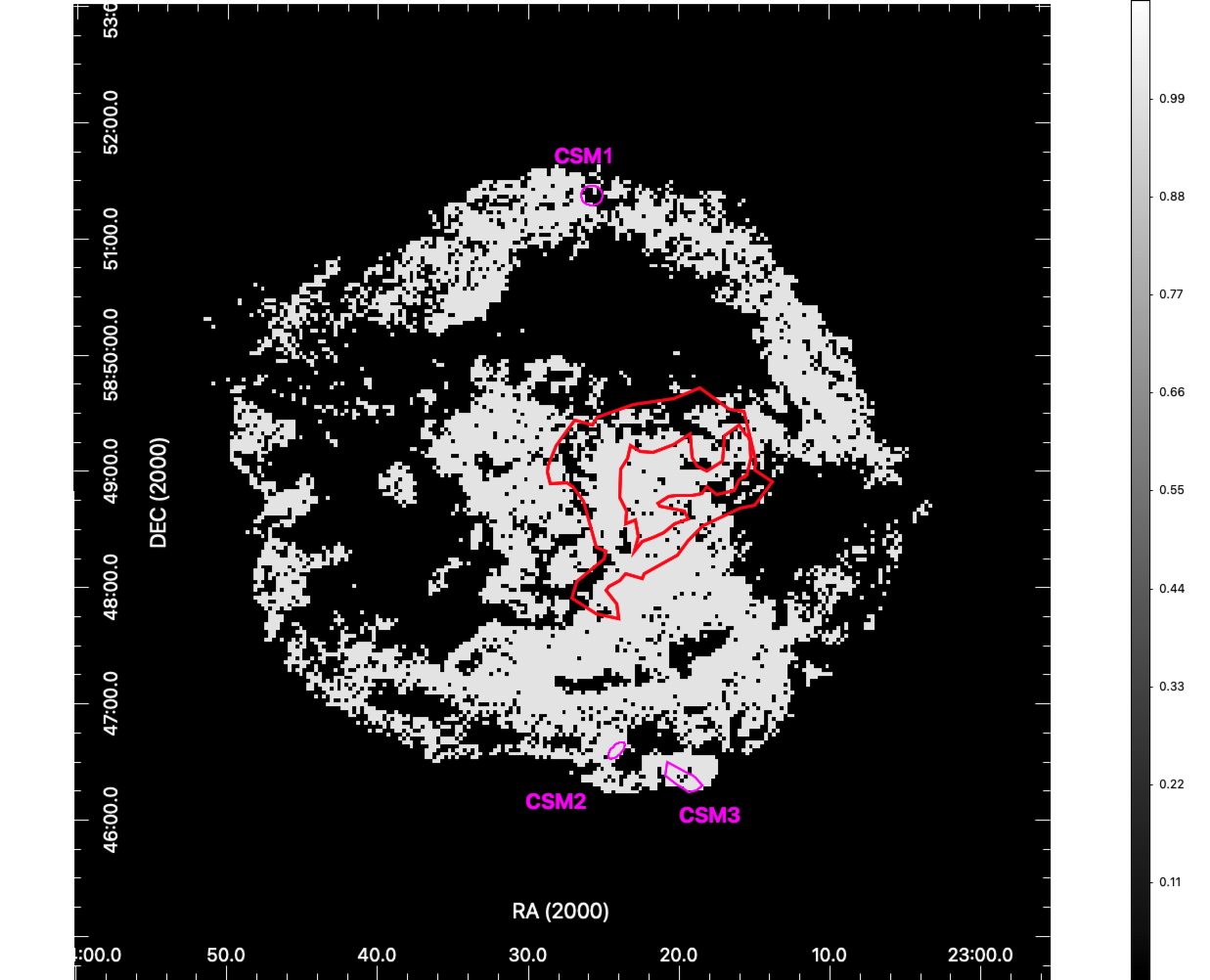}
}
\caption{Top panels: 
Scatterplot of PC scores for all pixels (black) and those corresponding to the GM region (red),  zooming in on where the majority of the points are located.
(See Fig.~\ref{fig:pcscatter_full} for the full range of PC scores.)
Bottom panels:
Mask created from the PC1 to PC8 scores using values corresponding to the red colored regions in the top panels.
The left panel corresponds to a lose selection and the right panel to a more stringent selection (see text).
\label{fig:pcscatter}
}
\end{figure*}

The PC scores 1--8 are shown in Fig.~\ref{fig:pcscores}. 
PC2 identifies mostly with the nonthermal X-ray emission, and PC3 and PC4 seem to identify with the (partially) iron-rich regions.
PC5 appears to correlate best with some filaments associated with the GM. However, PCA regularizes data as a set of orthogonal components,
whereas different physical components present in the X-ray spectra may not necessarily be orthogonal. Physical components may, therefore, be spread over
different PC scores. Scatter plots of PC scores for the total data set and just the GM region  (red) are shown in the top panels of Fig.~\ref{fig:pcscatter}.

We use these scatter plots to automatically select PC scores similar to the GM. We can identify all rebinned Chandra image pixels for which the spectra have similar PC properties as those of the GM. 
For this, we looked at the  density of GM points in scatter plots like the ones shown in Fig.~\ref{fig:pcscores}. 
In practice, we converted the scatter plots into two-dimensional histograms, with the option of doing this only for a selected
spatial region, such as the GM region (the red points in Fig.~\ref{fig:pcscatter}). 
In a pair-wise manner---for two scores---we then scan the individual score maps and determine whether the score values for each pixel correspond 
with a high enough point density in the scatter plot for the GM region. 
If this is indeed the case, we select this pixel in a mask image corresponding to a specific score, the mask image
containing only zeroes (not selected) and ones (selected). 
The end results is that for each score map we have a mask image, indicating locations with similar PCA scores as
the GM region.
The selection threshold density in the scatter plots is a tuneable parameter.
Fig.~\ref{fig:pcscatter} shows masks of all pixels where the PC1--8 properties are similar to the GM for two different selection thresholds. This mask consists of a multiplication of the individual masks corresponding to PC1 to PC8.
For the lenient threshold (Fig.~\ref{fig:pcscatter}, left), we see that the mask selects out all regions associated with the main shell, which is dominated by Si-rich ejecta emission. 
It also excludes the Fe-rich regions in the southeastern part (also reddish in Fig.~\ref{fig:casa_regions}, top). Instead, most of the outer regions, where we expect the shocked CSM plasma to reside,
 are selected. For the more stringent threshold  (Fig.~\ref{fig:pcscatter}, right), some parts of the outer regions are not selected,  which includes, interestingly, CSM1. 
 But the regions containing the arc of QSFs in the south (CSM2 and CSM3) are shown to be `` GM-like".
 This may be a hint that there are some intrinsic differences among the CSM properties throughout Cas A, either in composition, or in density, temperature and ionization age.
Note that the threshold used  here is rather severe as many areas within the GM-region itself are blacked out, perhaps due to contamination with ejecta  emission or nonthermal emission.

No matter the selection threshold, the PCA analysis identifies the spectral X-ray properties of the GM to be very dissimilar from the ejecta-rich shell, and to be much more similar to regions in the outskirts
of Cas A, where we expect the emission to come from shocked CSM. The dependency on threshold indicates some variations within both the GM-region as well as the outer regions of Cas A, which may be related
to differences in plasma properties.

\section{Discussion}

The GM is a striking, newly JWST/MIRI identified IR structure inside the main shell of Cas A. The JWST/MIRI data by itself already provides some clues that it is a structure related to the shocked CSM.
The X-ray analysis described here confirms a CSM origin. It shows 
that the X-ray spectra of GM regions are similar to the spectra of  CSM regions, although for all these spectra there is a variable level of contribution of X-ray synchrotron radiation.
A perhaps even more convincing argument that the GM X-ray spectra are CSM related is provided by the PCA, which shows that the spectral properties of the GM region also identifies the outer regions
of Cas A, for which the most likely origin is shocked CSM \citep[c.f. Fig.~5 in][]{hwang12}. 

From the X-ray spectral analysis a few commonalities, informing us on the nature of the GM plasma, emerge:
\begin{enumerate}
\itemsep 0 em
\item the ionization ages ($\tau$ or $n_{\rm e}t$) of both the GM and CSM are relatively low, $\tau_{\rm max}\approx 0.6$--$3.6\times 10^{11}~{\rm cm^{-3}s}$,
and falls within the range of the forward-shock-associated values reported by \citet{hwang12};
\item the GM spectra are all blueshifted,  corresponding to radial velocities of $\approx -1200$ to $-2900~ {\rm km\,s^{-1}}$, with a median velocity of $\approx -2300 ~ {\rm km\,s^{-1}}$;
\item some spectra require a synchrotron contribution, which dominates the spectrum of GM3; 
\item  the CSM spectra are rather similar to the GM spectra in that the 
silicon/sulfur line emission and/or iron-line emission are much less prominent (i.e. have lower equivalent widths) than for typical spectra from Cas A \citep[c.f. spectra in][]{hwang12};
\item the abundance pattern for the GM spectra is  similar to those of the CSM spectra.
\end{enumerate}

In particular, points 1) and 4) reinforce the idea that the GM spectra are indeed  CSM-related. The presence of a strong synchrotron component in some spectra (in particular GM3)
suggests that these filaments may be close to the forward shock, as the rapid synchrotron cooling times of the electrons with $\gtrsim 10$~TeV only allows for X-ray synchrotron radiation within
$10^{17}$~cm of the shock \citep[e.g.][]{vink03a}. But chance alignment with either forward-shock or reverse-shock-related filaments cannot be excluded.

The abundance pattern or both GM and CSM regions does show relatively low abundances, but are enhanced with respect to solar abundances.
Optical studies indicate that the CSM of Cas A is enhanced in He and N, which will affect the thermal continuum,
and hence the derived abundances of the other elements.
In particular, the oxygen abundance both for the GM and the CSM spectra is relatively high.
It is not clear whether the CSM of Cas A is indeed enhanced in oxygen or whether some systematics are involved. We note that oxygen is nearly completely ionized, and the low energy part of the spectrum with 
oxygen lines is heavily affected by absorption. Both effects may leave the measured abundance vulnerable to systematic effects.

The median $\tau_{\rm max}$ value for the GM regions is $\tau_{\rm max}=1.17
\times 10^{11}~{\rm cm^{-3}s}$, and the mean value $\tau_{\rm max}=1.8
\times 10^{11}~{\rm cm^{-3}s}$. Taking $1.5\times 10^{11}~{\rm cm^{-3}s}$
to be typical for the GM we find that the typical density 
is $n_{\rm e,1}\approx \tau_{\rm max}/t= 47~{\rm cm^{-3}}(t/100~{\rm yr})(\tau_{\rm max}/1.5\times 10^{11})$, with $t$ the time since the plasma
was shocked.
This 
corresponds
to a pre-shock density---i.e. a factor 4 lower (due to the compression factor)---of  $n_{\rm e,0}\approx 12/(t/100~{\rm yr})(\tau_{\rm max}/1.5\times 10^{11})~{\rm cm^{-3}}$. 
This value for the pre-shock density is a factor 5--10 higher than previous estimates \citep{vink04a,lee14} but could be marginally consistent if one takes plasma ages closer to the maximum age of $\sim 350$~yr.

The $\tau_{\rm max}$ for the CSM regions is still relatively
high, but consistent with the CSM regions analyzed by \citet{hwang12}.
It is interesting that the regions associated with the QSFs (CSM2 and CSM3)
have $\tau_{\rm max}$ values comparable to the GM regions,
whereas CSM1 has a low value of $\tau_{\rm max}\approx 0.6\times 10^{11}~{\rm cm^{-3}s}$. The latter would correspond to a pre-shock density
of $n_{\rm e,0}\approx 5/(t/100~{\rm yr})(\tau_{\rm max}/0.6\times 10^{11})~{\rm cm^{-3}}$---on the high side, but more consistent with the 1--2~${\rm cm^{-3}}$ pre-shock density estimates of \citet{vink04a,lee14}.

The blueshifts of the GM spectra indicate that the GM lies on the near-side of Cas A, and, as shown in Fig.~\ref{fig:cornerplots}, are statistically  significant. 
Interestingly, the QSF regions are also on average blueshifted, with  optically determined Doppler shifts of approximately -600--100~${\rm km\,s^{-1}}$ \citep{vandenbergh71,alarie14}. 
Another hint that there may be a connection between the  QSFs and the GM \citep[see also][]{delooze24}. 

The regions CSM2 and CSM3 analyzed here, are associated with the QSFs. 
The spectrum of CSM2 shows a blueshift of $\approx 1450~{\rm km\,s^{-1}}$, which is surprisingly large
for a region that is close to the edge of Cas A.
CSM2 is located in a region where the measured X-ray proper motions indicate that the shock velocity changes from a relatively high value of $\sim 6200~{\rm km\,s^{-1}}$ to
$\sim 4000~{\rm km\,s^{-1}}$ \citep{vink22a}. So a high shock velocity around CSM2 may indeed be  possible,
leading to a relatively large residual radial velocity. But this needs to be further investigated.

The typical  GM's radial velocity of $|v_{\rm rad}|\approx 2300 = v_{\rm plasma}\cos i~{\rm km\ s^{-1}}$ corresponds to a shock velocity of $V_{\rm s}\approx 3100/\cos i~{\rm km\ s^{-1}}$,  given that the plasma downstream
of a strong shock running into a monatomic gas has $v_{\rm plasma}=3/4 V_{\rm s}$ in the frame of the observer. 
The factor $\cos i$ takes into account projection effects, with $i$ the inclination angle. Since the GM lies close to the center,
the projection effects should be small ---$i\approx 0$---30\deg---which suggests that the GM is associated with parts of the CSM where the shock front has decelerated to $V_{\rm s}\sim 3500~{\rm km\ s^{-1}}$.
X-ray proper motion studies of Cas A show that the typical  shock speeds are 5800~km\,s$^{-1}$ \citep{vink22a}. But a value as low as 4038~km\ s$^{-1}$ is reported for a position angle of 190\deg. Interestingly, this corresponds
to the region containing the arc of QSFs, for which a higher than average density is likely.
Since  the shock velocities for a given postshock pressure are proportional to  $\rho_0 V_{\rm s}^2$, we expect that the lower inferred shock velocities are the result of a pre-shock density that is
$V_{\rm s,GM}^2/\overline{V_{\rm s}}^2\approx 3$ higher than the average CSM density near the current shock front.
So both the best-fit $\tau_{\rm max}$  values, as well as the relatively low radial velocities of $\approx 2300~{\rm km\ s^{-1}}$ suggest that the filaments in the GM stand out due to a higher than average density.

Some of the X-ray analysis presented here goes beyond the main topic of this paper, i.e. the nature of the GM. For example, the PCA provides a yet unexplored way to identify which  parts of Cas A consist mostly
of shocked CSM and which parts consist of shocked ejecta. It is interesting that 
most of the  region interior to the main shell---not just the GM---seems to be spectroscopically similar to the shocked CSM. This suggests a lack of ejecta-related emission from the central regions of Cas A.
According to the PCA CSM1 appears to have  different X-ray properties than the GM regions and CSM2 and CSM3, which may reflect some 
variation in thermal X-ray emission from the shocked CSM.
This may be related to the variation in IR dust grain emission properties between the northern CSM on the one hand, and the QSF and GM-related emission on the other hand,
which is due to a higher carbon-over-silicate abundance in the QSF and GM regions \citep{delooze24}.

Another noteworthy result of the PCA is that
the main shell for position angles
between roughly 170\deg and 200\deg seems to have some CSM-like properties, except for two narrow gaps.  This is a region in which the ejecta shell is relatively narrow in X-rays (Fig.~\ref{fig:casa_regions}, top) and in
the optical, the ejecta-related fast moving knots only started appearing since the 1970s \citep[see Fig.~1 in][]{patnaude14}. Some of the IR emission properties in that region seem to be similar to the GM emission \citep{milisavljevic24,delooze24}. In addition,  this region is also relatively bright in [Fe II] $1.644~{\rm \mu m}$, which it has in common with the arc of QSFs further to the south \citep[][their Fig.~12]{lee17}.
\cite{lee17} attribute this to a prominent CSM component associated with the contact discontinuity in this part of the shell.

\section{Conclusion}

We analyzed the  X-ray properties of the JWST-identified ``Green Monster" (GM) region in order to study the nature of this new structure. For the X-ray investigation, we used a  spectral analysis employing
an absorbed {\tt vpshock} plus power-law model, the latter to account for nonthermal X-ray emission. For optimizing the models we used  a Bayesian analysis scheme.
In addition, we characterized the X-ray spectral properties of the GM in comparison to the rest of Cas A using PCA.

Both types of  X-ray analysis suggest that the GM is a structure that has similar spectral characteristics as the outer regions of Cas A, which consist most likely of shocked  CSM.
For the PCA, this conclusion is based on the fact that selecting PC scores associated with the GM also select for the outer regions of Cas A, and exclude the Si-rich and Fe-rich ejecta components.
Spectroscopically we show that GM spectra are relatively weak in Si XIII emission---and line emission from other intermediate mass elements---and are well characterized by abundances that are close to
solar abundances, and quite similar to other CSM-related spectra. 

The analysis presented here supports the conclusion of \citet{delooze24}, based on JWST data alone, that the GM monster structure corresponds to a structure in the CSM of Cas A, rather than to a  ejecta-related structure.

In addition, the X-ray analysis shows that the GM spectra are blueshifted with a median radial velocity of $\approx -2300~{\rm km\ s^{-1}}$, suggesting that the GM structure is on the near-side of Cas A, and projected onto the interior. 
The radial velocity is relatively low,  given the average shock speed of Cas A of 5800~km\,s$^{-1}$,
which could be explained if the pre-shock CSM corresponding to the GM was a factor 2--3 denser than the average CSM density.

The infrared and X-ray properties of the GM provide yet another piece of the puzzle regarding the CSM of Cas A. This will help  to reconstruct the mass-loss properties of the progenitor star,
and how it ended up to become the Type IIb supernova identified from light echo spectra \citep{krause08,rest11}.

\begin{acknowledgements}
The authors thank the Lorentz Center Leiden and the main scientific organizer Maria Arias for organizing the workshop ``Supernova remnants in complex environments", where most of the authors were present.
We thank Amaël Ellien and Emanuele Greco for helping M.A. set up the Bayesian spectral fitting procedures, and we thank Allessandra Mercuri for her help in understanding the X-ray properties
of the shocked CSM in Cas A.
M.A. is supported by the research program Athena with project No. 184.034.002, which is (partially) financed by the Dutch Research Council (NWO). J.V. is supported by funding from the European Unions Horizon 2020 research and innovation program under grant agreement No. 101004131 (SHARP).
D.M. acknowledges NSF support from grants PHY- 2209451 and AST-2206532. 
\end{acknowledgements}


\begin{thebibliography}{}
\expandafter\ifx\csname natexlab\endcsname\relax\def\natexlab#1{#1}\fi
\providecommand{\url}[1]{\href{#1}{#1}}
\providecommand{\dodoi}[1]{doi:~\href{http://doi.org/#1}{\nolinkurl{#1}}}
\providecommand{\doeprint}[1]{\href{http://ascl.net/#1}{\nolinkurl{http://ascl.net/#1}}}
\providecommand{\doarXiv}[1]{\href{https://arxiv.org/abs/#1}{\nolinkurl{https://arxiv.org/abs/#1}}}

\bibitem[{{Alarie} {et~al.}(2014){Alarie}, {Bilodeau}, \& {Drissen}}]{alarie14}
{Alarie}, A., {Bilodeau}, A., \& {Drissen}, L. 2014, \mnras, 441, 2996,
  \dodoi{10.1093/mnras/stu774}

\bibitem[{{Arnaud}(1996)}]{arnaud96}
{Arnaud}, K.~A. 1996, Astronomical Society of the Pacific Conference Series,
  Vol. 101, {XSPEC: The First Ten Years}, 17

\bibitem[{{Ashton} {et~al.}(2022){Ashton}, {Bernstein}, {Buchner}, {Chen},
  {Cs{\'a}nyi}, {Fowlie}, {Feroz}, {Griffiths}, {Handley}, {Habeck}, {Higson},
  {Hobson}, {Lasenby}, {Parkinson}, {P{\'a}rtay}, {Pitkin}, {Schneider},
  {Speagle}, {South}, {Veitch}, {Wacker}, {Wales}, \& {Yallup}}]{ashton22}
{Ashton}, G., {Bernstein}, N., {Buchner}, J., {et~al.} 2022, Nature Reviews
  Methods Primers, 2, 39, \dodoi{10.1038/s43586-022-00121-x}

\bibitem[{{Baade} \& {Minkowski}(1954)}]{baade54}
{Baade}, W., \& {Minkowski}, R. 1954, \apj, 119, 206, \dodoi{10.1086/145812}

\bibitem[{{Blackburn}(1995)}]{ftools}
{Blackburn}, J.~K. 1995, in Astronomical Society of the Pacific Conference
  Series, Vol.~77, Astronomical Data Analysis Software and Systems IV, ed.
  R.~A. {Shaw}, H.~E. {Payne}, \& J.~J.~E. {Hayes}, 367

\bibitem[{{Borkowski} {et~al.}(2001){Borkowski}, {Rho}, {Reynolds}, \&
  {Dyer}}]{borkowski01}
{Borkowski}, K.~J., {Rho}, J., {Reynolds}, S.~P., \& {Dyer}, K.~K. 2001, \apj,
  550, 334, \dodoi{10.1086/319716}

\bibitem[{{Braun}(1987)}]{braun87}
{Braun}, R. , {Gull}, S. F., {Perley}, R. A., 1987, \nat, 327, 395, \dodoi{10.1038/327395a0}

\bibitem[{{Buchner}(2021)}]{buchner21}
{Buchner}, J. 2021, The Journal of Open Source Software, 6, 3001,
  \dodoi{10.21105/joss.03001}

\bibitem[{{Buchner}(2023)}]{buchner23}
---. 2023, Statistics Surveys, 17, 169, \dodoi{10.1214/23-SS144}

\bibitem[{{Buchner} {et~al.}(2014){Buchner}, {Georgakakis}, {Nandra}, {Hsu},
  {Rangel}, {Brightman}, {Merloni}, {Salvato}, {Donley}, \&
  {Kocevski}}]{buchner14}
{Buchner}, J., {Georgakakis}, A., {Nandra}, K., {et~al.} 2014, \aap, 564, A125,
  \dodoi{10.1051/0004-6361/201322971}

\bibitem[{{De Looze} {et~al.}(2024)}]{delooze24}
{De Looze}, I., {et~al.} 2024, \apj\ (in preparation)

\bibitem[{{Eadie} {et~al.}(2023){Eadie}, {Speagle}, {Cisewski-Kehe},
  {Foreman-Mackey}, {Huppenkothen}, {Jones}, {Springford}, \& {Tak}}]{eadie23}
{Eadie}, G.~M., {Speagle}, J.~S., {Cisewski-Kehe}, J., {et~al.} 2023, arXiv
  e-prints, arXiv:2302.04703, \dodoi{10.48550/arXiv.2302.04703}

\bibitem[{{Ellien} {et~al.}(2023){Ellien}, {Greco}, \& {Vink}}]{ellien23}
{Ellien}, A., {Greco}, E., \& {Vink}, J. 2023, \apj, 951, 103,
  \dodoi{10.3847/1538-4357/accc85}

\bibitem[{{Gotthelf} {et~al.}(2001){Gotthelf}, {Koralesky}, {Rudnick}, {Jones},
  {Hwang}, \& {Petre}}]{gotthelf01a}
{Gotthelf}, E.~V., {Koralesky}, B., {Rudnick}, L., {et~al.} 2001, \apjl, 552,
  L39, \dodoi{10.1086/320250}

\bibitem[{{Helder} \& {Vink}(2008)}]{helder08}
{Helder}, E.~A., \& {Vink}, J. 2008, \apj, 686, 1094, \dodoi{10.1086/591242}

\bibitem[{{Hughes} {et~al.}(2000){Hughes}, {Rakowski}, {Burrows}, \&
  {Slane}}]{hughes00a}
{Hughes}, J.~P., {Rakowski}, C.~E., {Burrows}, D.~N., \& {Slane}, P.~O. 2000,
  \apjl, 528, L109, \dodoi{10.1086/312438}

\bibitem[{{Hwang} \& {Laming}(2012)}]{hwang12}
{Hwang}, U., \& {Laming}, J.~M. 2012, \apj, 746, 130,
  \dodoi{10.1088/0004-637X/746/2/130}

\bibitem[{{Hwang} {et~al.}(2004)}]{hwang04}
{Hwang}, U., {et~al.} 2004, \apjl, 615, L117, \dodoi{10.1086/426186}

\bibitem[{{Kaastra} \& {Bleeker}(2016)}]{Kaastra16}
{Kaastra}, J.~S., \& {Bleeker}, J.~A.~M. 2016, \aap, 587, A151,
  \dodoi{10.1051/0004-6361/201527395}

\bibitem[{{Koo} {et~al.}(2023){Koo}, {Kim}, {Yoon}, \& {Raymond}}]{koo23}
{Koo}, B.-C., {Kim}, D., {Yoon}, S.-C., \& {Raymond}, J.~C. 2023, \apj, 945,
  158, \dodoi{10.3847/1538-4357/acb7e7}

\bibitem[{{Koo} {et~al.}(2020){Koo}, {Kim}, {Oh}, {Raymond}, {Yoon}, {Lee}, \&
  {Jaffe}}]{koo20}
{Koo}, B.-C., {Kim}, H.-J., {Oh}, H., {et~al.} 2020, Nature Astronomy, 4, 584,
  \dodoi{10.1038/s41550-019-0996-4}

\bibitem[{{Krause} {et~al.}(2008){Krause}, {Birkmann}, {Usuda}, {Hattori},
  {Goto}, {Rieke}, \& {Misselt}}]{krause08}
{Krause}, O., {Birkmann}, S.~M., {Usuda}, T., {et~al.} 2008, Science, 320,
  1195, \dodoi{10.1126/science.1155788}

\bibitem[{{Lee} {et~al.}(2014){Lee}, {Park}, {Hughes}, \& {Slane}}]{lee14}
{Lee}, J.-J., {Park}, S., {Hughes}, J.~P., \& {Slane}, P.~O. 2014, \apj, 789,
  7, \dodoi{10.1088/0004-637X/789/1/7}

\bibitem[{{Lee} {et~al.}(2017){Lee}, {Koo}, {Moon}, {Burton}, \& {Lee}}]{lee17}
{Lee}, Y.-H., {Koo}, B.-C., {Moon}, D.-S., {Burton}, M.~G., \& {Lee}, J.-J.
  2017, \apj, 837, 118, \dodoi{10.3847/1538-4357/aa60c0}

\bibitem[{{Lodders} {et~al.}(2009){Lodders}, {Palme}, \& {Gail}}]{lodders09}
{Lodders}, K., {Palme}, H., \& {Gail}, H.~P. 2009, Landolt B\"ornstein, 4B,
  712, \dodoi{10.1007/978-3-540-88055-4_34}

\bibitem[{{Milisavljevic} {et~al.}(2024){Milisavljevic}, {Temim}, {De Looze},
  {Dickinson}, {Laming}, {et~al.}}]{milisavljevic24}
{Milisavljevic}, D., {Temim}, T., {De Looze}, I., {et~al.} 2024, arXiv
  e-prints, arXiv:2401.02477, \dodoi{10.48550/arXiv.2401.02477}

\bibitem[{{Patnaude} \& {Fesen}(2014)}]{patnaude14}
{Patnaude}, D.~J., \& {Fesen}, R.~A. 2014, \apj, 789, 138,
  \dodoi{10.1088/0004-637X/789/2/138}

\bibitem[{{Patnaude} {et~al.}(2011){Patnaude}, {Vink}, {Laming}, \&
  {Fesen}}]{patnaude11}
{Patnaude}, D.~J., {Vink}, J., {Laming}, J.~M., \& {Fesen}, R.~A. 2011, \apjl,
  729, L28+, \dodoi{10.1088/2041-8205/729/2/L28}

\bibitem[{{Rest} {et~al.}(2011){Rest}, {Foley}, {Sinnott}, {Welch}, {Badenes},
  {Filippenko}, {Bergmann}, {Bhatti}, {Blondin}, {Challis}, {Damke}, {Finley},
  {Huber}, {Kasen}, {Kirshner}, {Matheson}, {Mazzali}, {Minniti}, {Nakajima},
  {Narayan}, {Olsen}, {Sauer}, {Smith}, \& {Suntzeff}}]{rest11}
{Rest}, A., {Foley}, R.~J., {Sinnott}, B., {et~al.} 2011, \apj, 732, 3,
  \dodoi{10.1088/0004-637X/732/1/3}

\bibitem[{{Uchiyama} \& {Aharonian}(2008)}]{uchiyama08}
{Uchiyama}, Y., \& {Aharonian}, F.~A. 2008, \apjl, 677, L105,
  \dodoi{10.1086/588190}

\bibitem[{{van den Bergh}(1971)}]{vandenbergh71}
{van den Bergh}, S. 1971, \apj, 165, 457, \dodoi{10.1086/150913}

\bibitem[{{Vink}(2004)}]{vink04a}
{Vink}, J. 2004, New Astronomy Review, 48, 61,
  \dodoi{10.1016/j.newar.2003.11.008}

\bibitem[{{Vink} \& {Laming}(2003)}]{vink03a}
{Vink}, J., \& {Laming}, J.~M. 2003, \apj, 584, 758, \dodoi{10.1086/345832}

\bibitem[{{Vink} {et~al.}(2022){Vink}, {Patnaude}, \& {Castro}}]{vink22a}
{Vink}, J., {Patnaude}, D.~J., \& {Castro}, D. 2022, \apj, 929, 57,
  \dodoi{10.3847/1538-4357/ac590f}

\bibitem[{{Warren}(2006)}]{warren06}
{Warren}, J.~S. 2006, PhD thesis, Rutgers University, New Jersey

\bibitem[{{Warren} \& {Hughes}(2004)}]{warren04}
{Warren}, J.~S., \& {Hughes}, J.~P. 2004, \apj, 608, 261

\bibitem[{{Wilms} {et~al.}(2000){Wilms}, {Allen}, \& {McCray}}]{wilms00}
{Wilms}, J., {Allen}, A., \& {McCray}, R. 2000, \apj, 542, 914,
  \dodoi{10.1086/317016}

\end{thebibliography}

\appendix

\renewcommand{\thefigure}{A\arabic{figure}}
\setcounter{figure}{0}
\renewcommand{\thetable}{A\arabic{table}}
\setcounter{table}{0}

\section*{Supplementary material}

\begin{figure*}[hbp]
\includegraphics[trim=0 0 120 0,clip=true,width=0.96\textwidth]{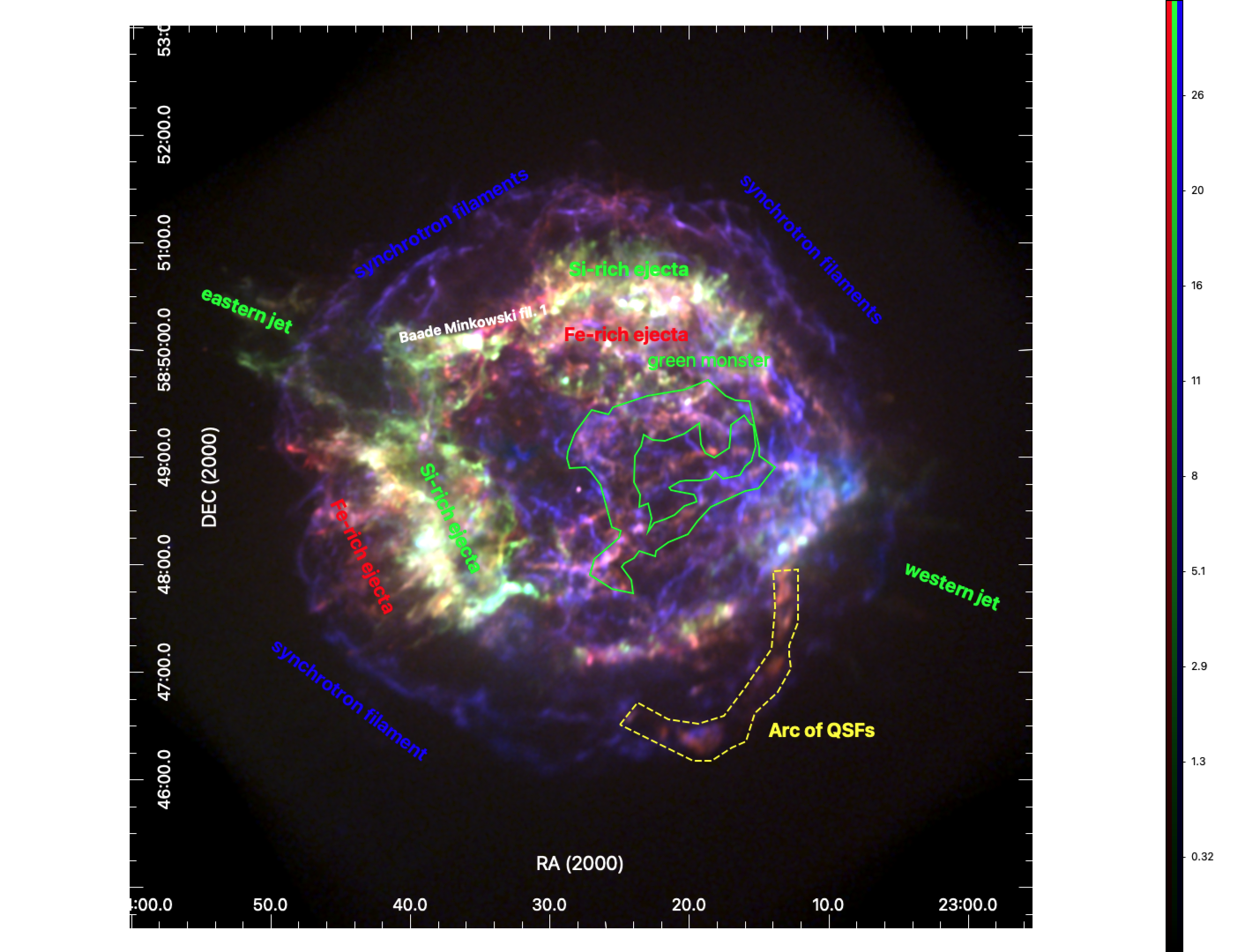}
\caption{
Same as Fig.~\ref{fig:casa_regions}, but now with rough indications of important regions mentioned in the text.
In general, all Si-rich ejecta show up here in green. Fe-rich ejecta show up as shades of red, but so are regions relatively abundant in Ne IX/X and Mg XI/XII.
Bright X-ray synchrotron filaments show up in shades of purple. 
\label{fig:casa_legend}
}
\end{figure*}

\begin{figure*}

  \centerline{
    \includegraphics[trim=0 0 0 0,clip=true,width=0.33\textwidth]{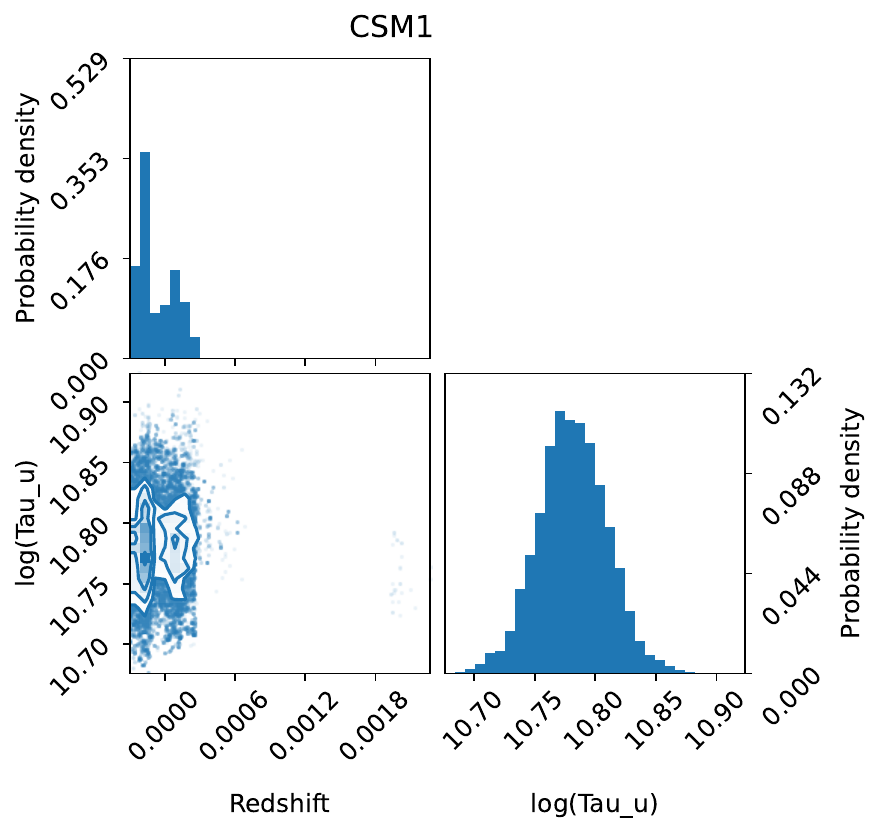}
    \includegraphics[width=0.33\textwidth]{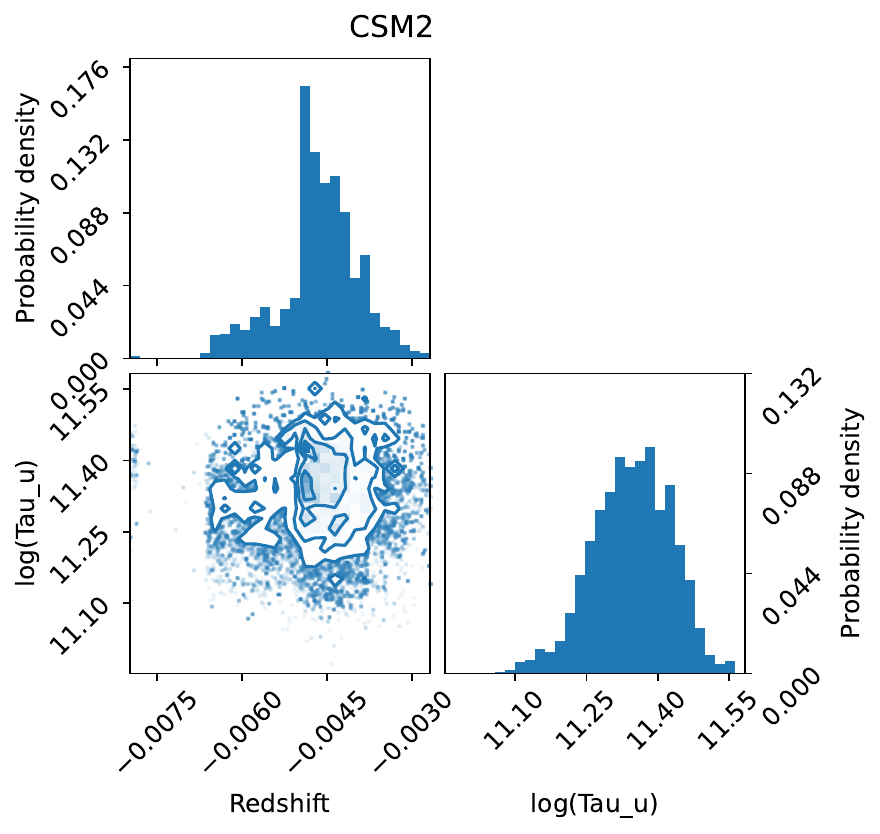}
    \includegraphics[width=0.33\textwidth]{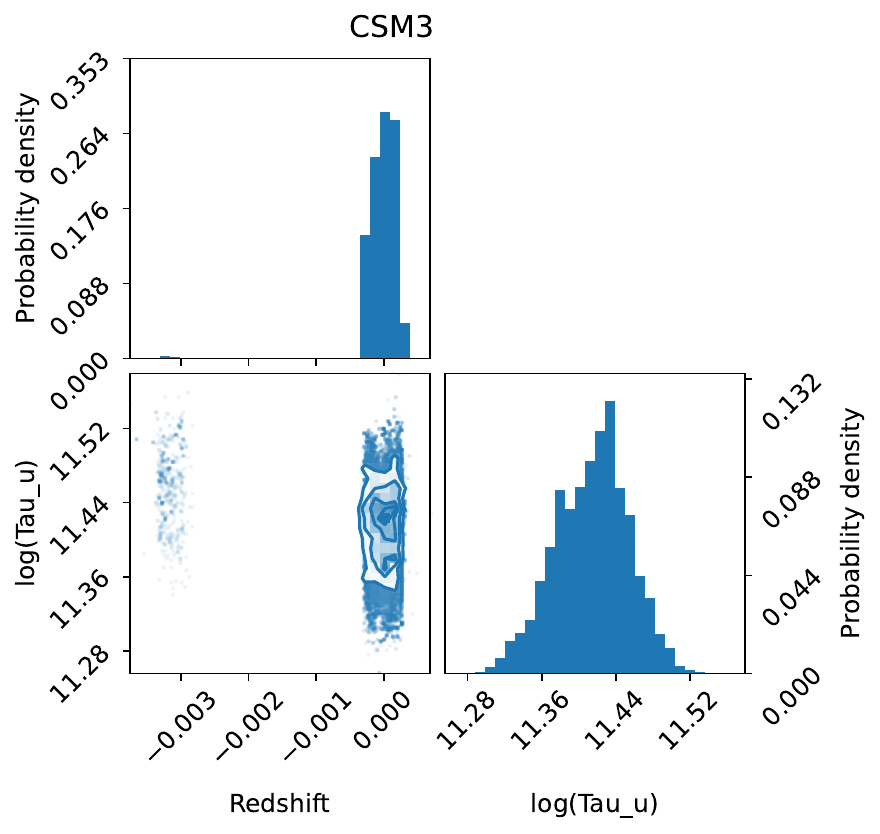}
  }
  \centerline{
    \includegraphics[width=0.33\textwidth]{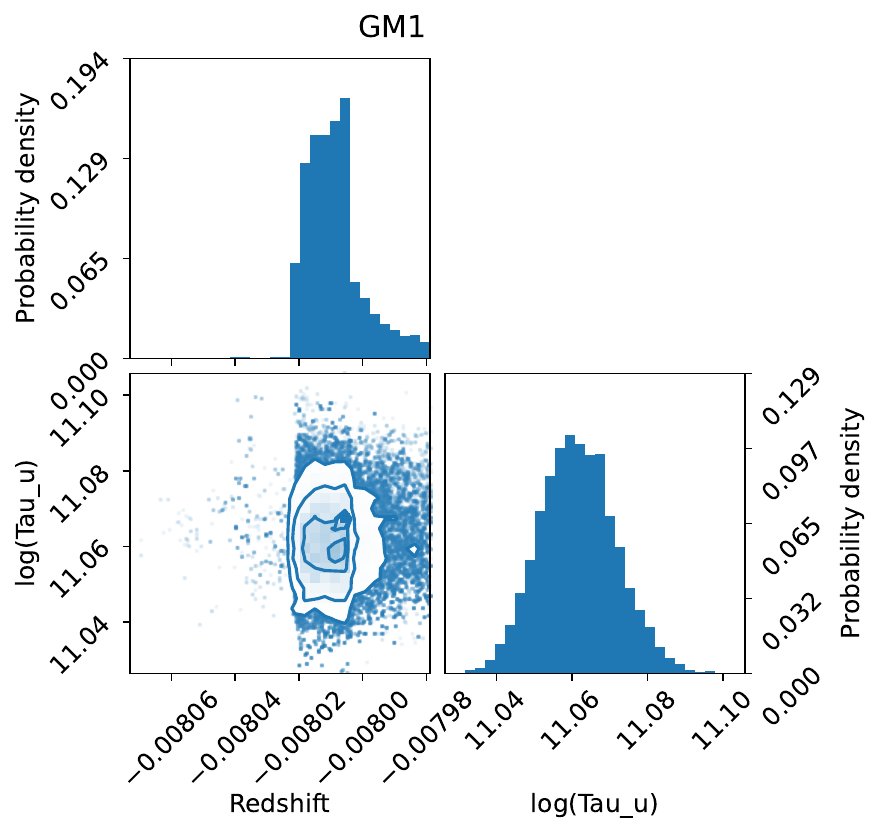}
    \includegraphics[width=0.33\textwidth]{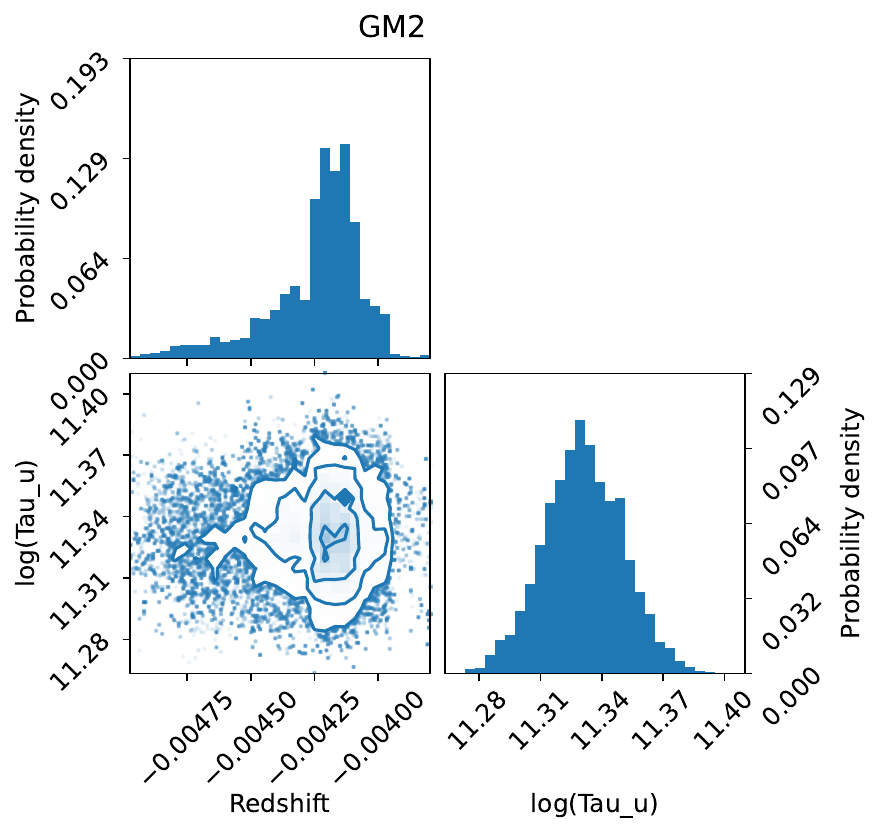}
    \includegraphics[width=0.33\textwidth]{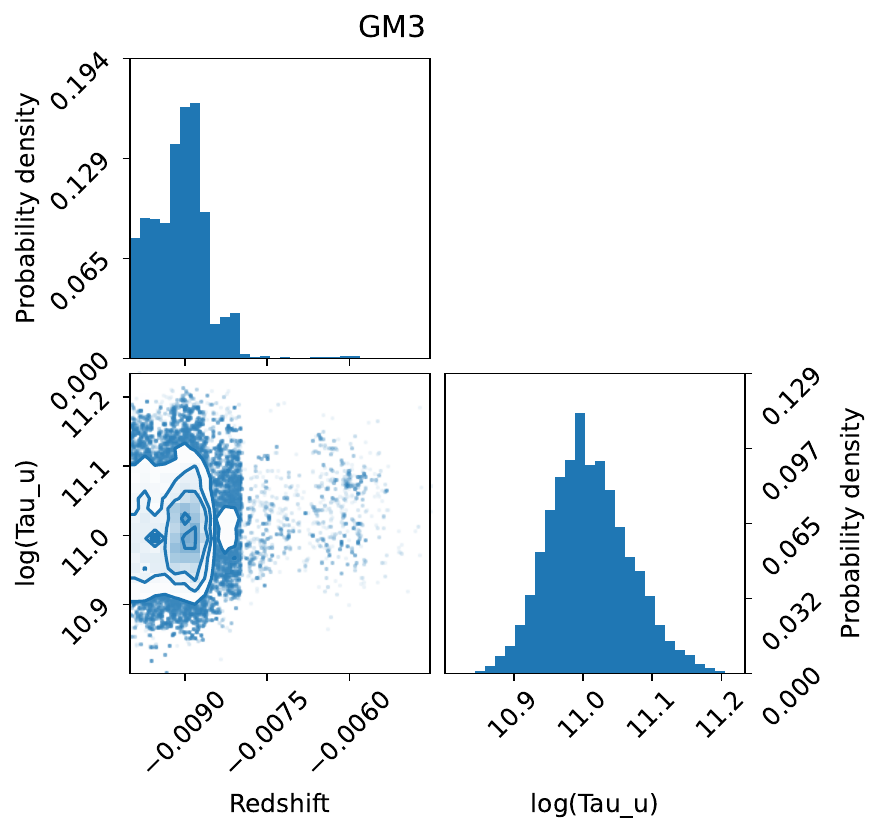}
  }
  \centerline{
    \includegraphics[width=0.33\textwidth]{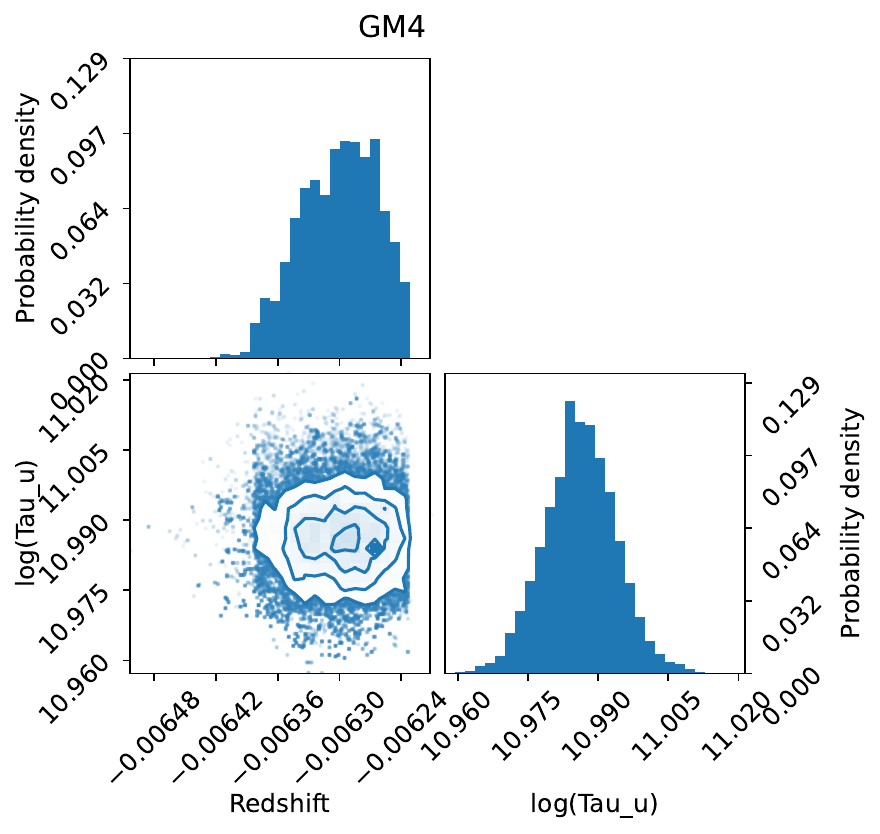}
    \includegraphics[width=0.33\textwidth]{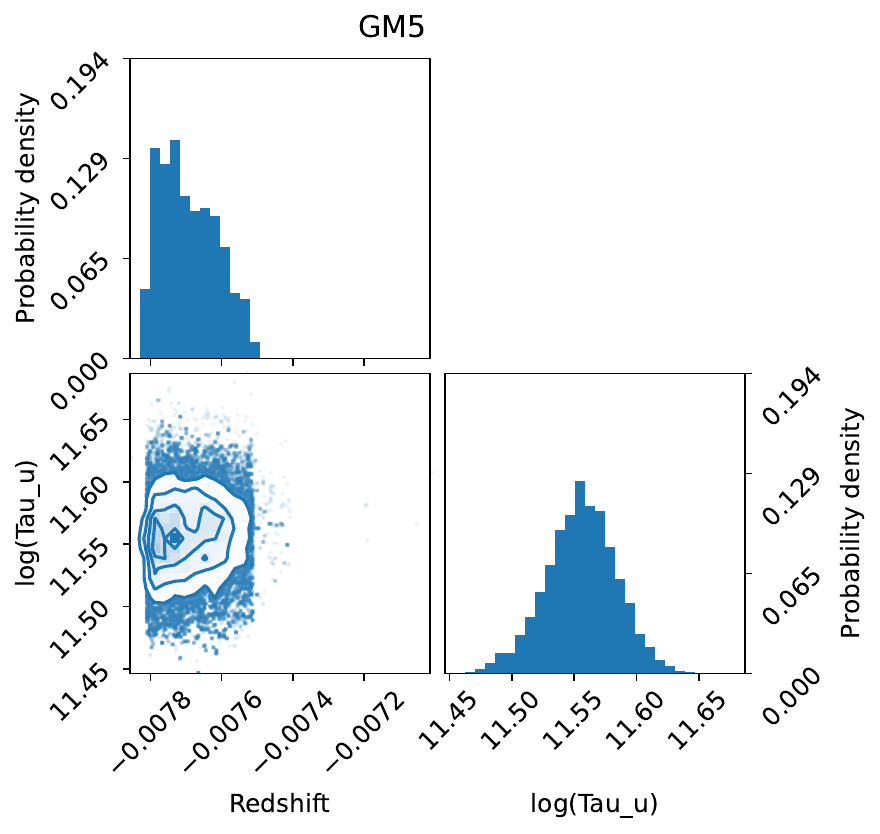}
  }

  \caption{Correlation plots of the Doppler shifts versus  $\tau_{\rm max}$ for
    all GM and CSM regions. The histograms are normalized to a sum of one to form a probability density distribution.
    The GM analysis is based on  155052 nested sampling iterations and for the CSM analysis there were 113453 nested sampling iterations.
    \label{fig:cornerplots}
  }
\end{figure*}

\begin{figure*}[hbp]

\centerline{
\includegraphics[trim=0 0 0 0,clip=true,width=0.25\textwidth]{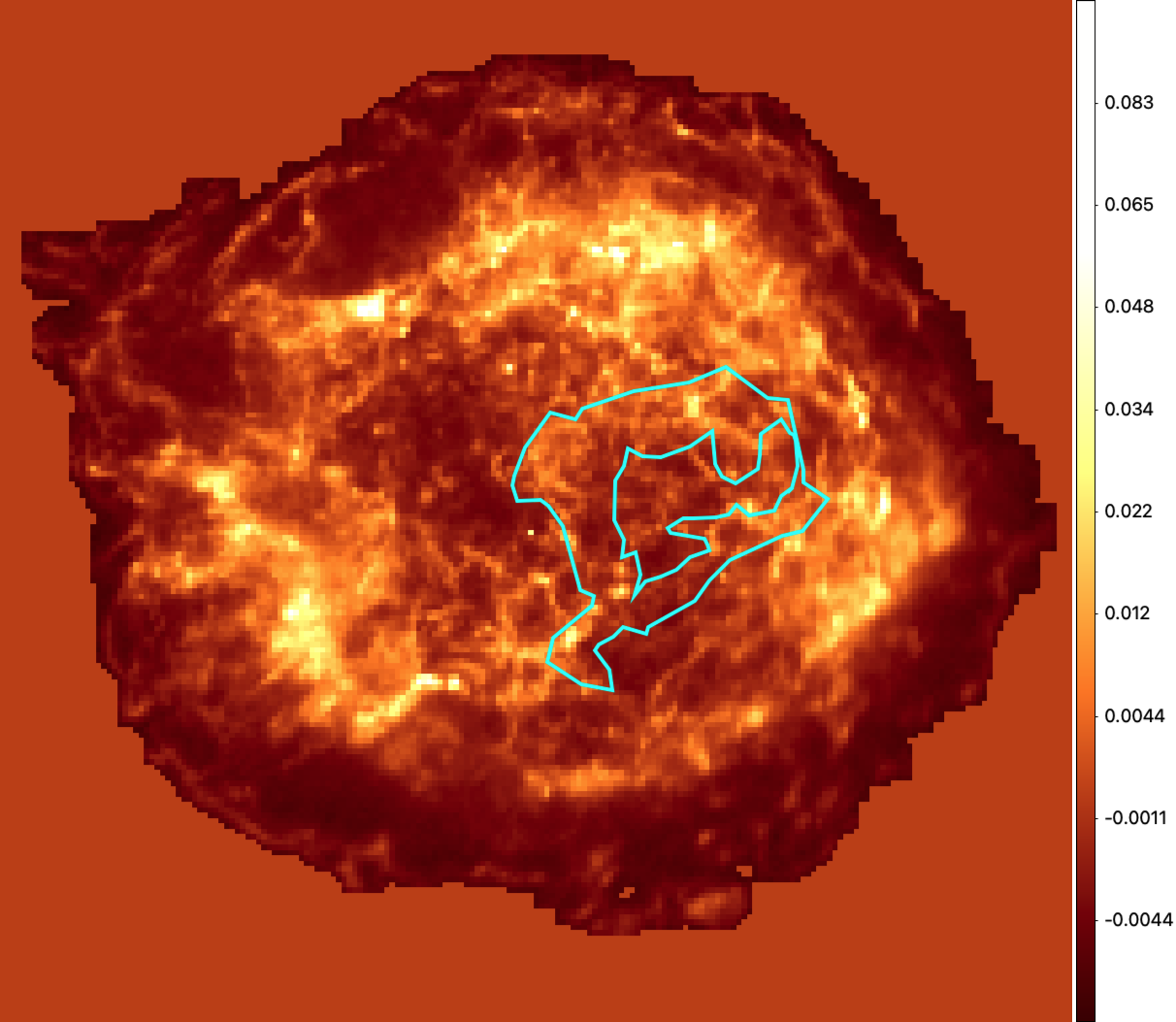}
\includegraphics[trim=0 0 0 0,clip=true,width=0.25\textwidth]{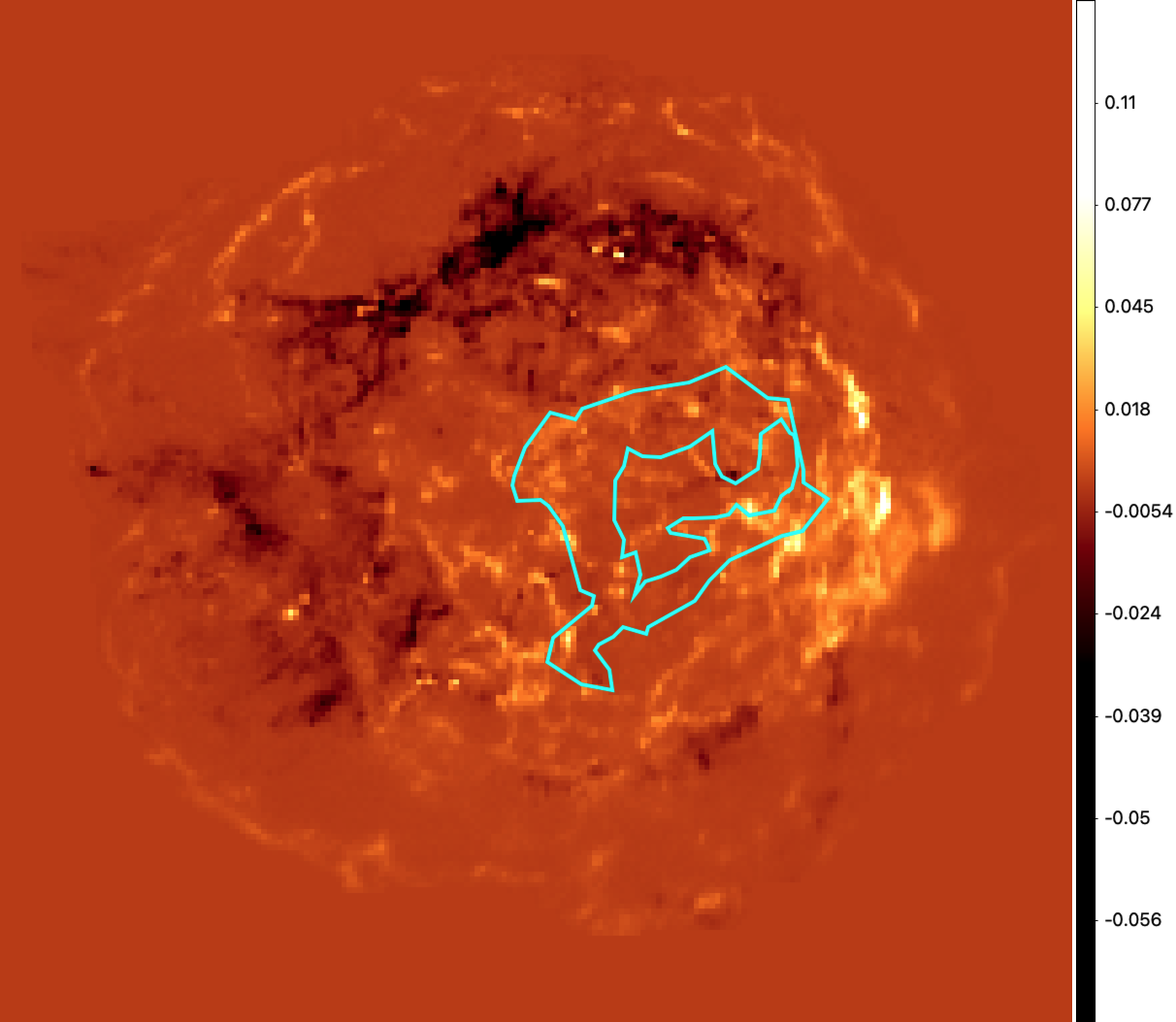}
\includegraphics[trim=0 0 0 0,clip=true,width=0.25\textwidth]{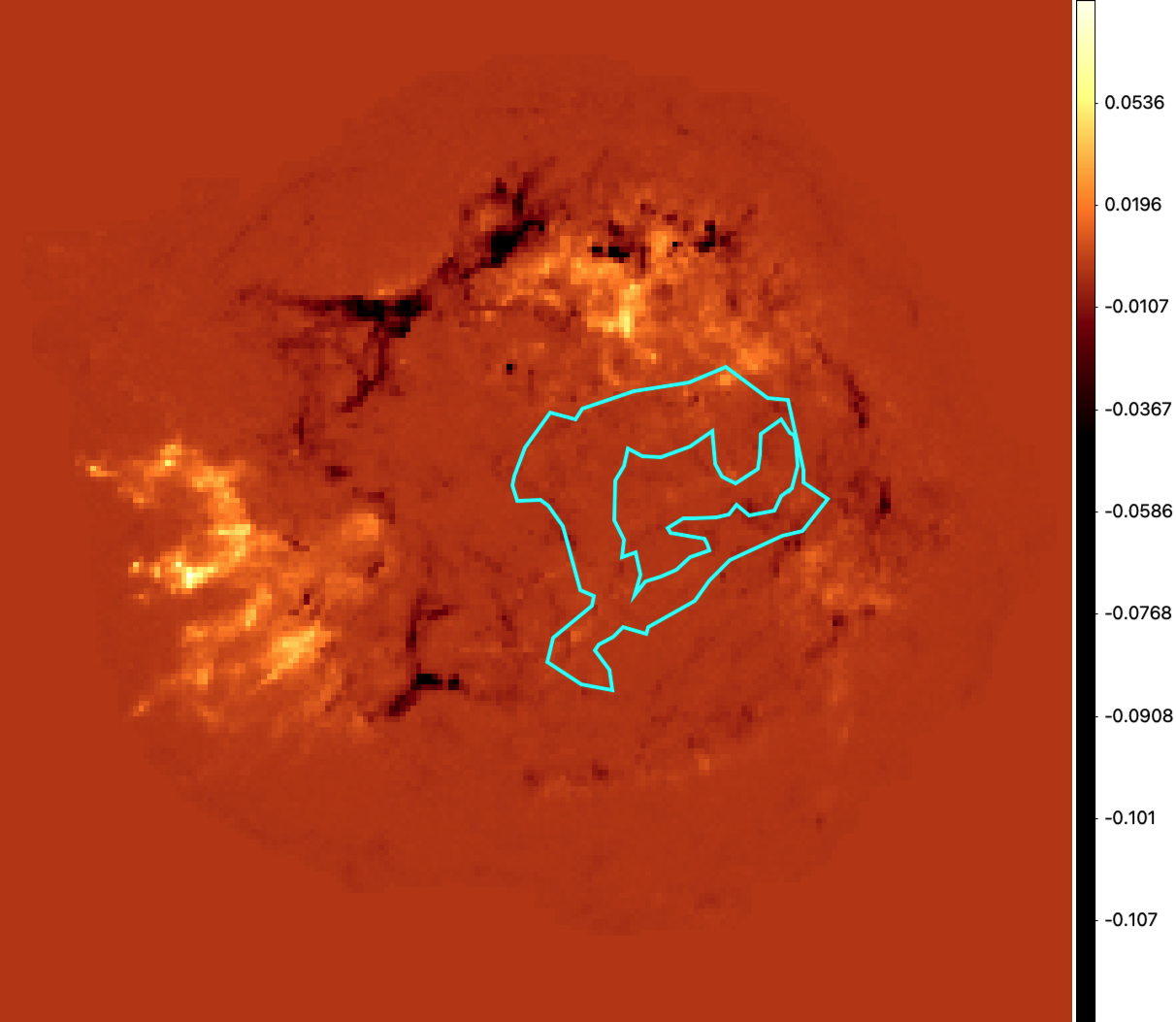}
\includegraphics[trim=0 0 0 0,clip=true,width=0.25\textwidth]{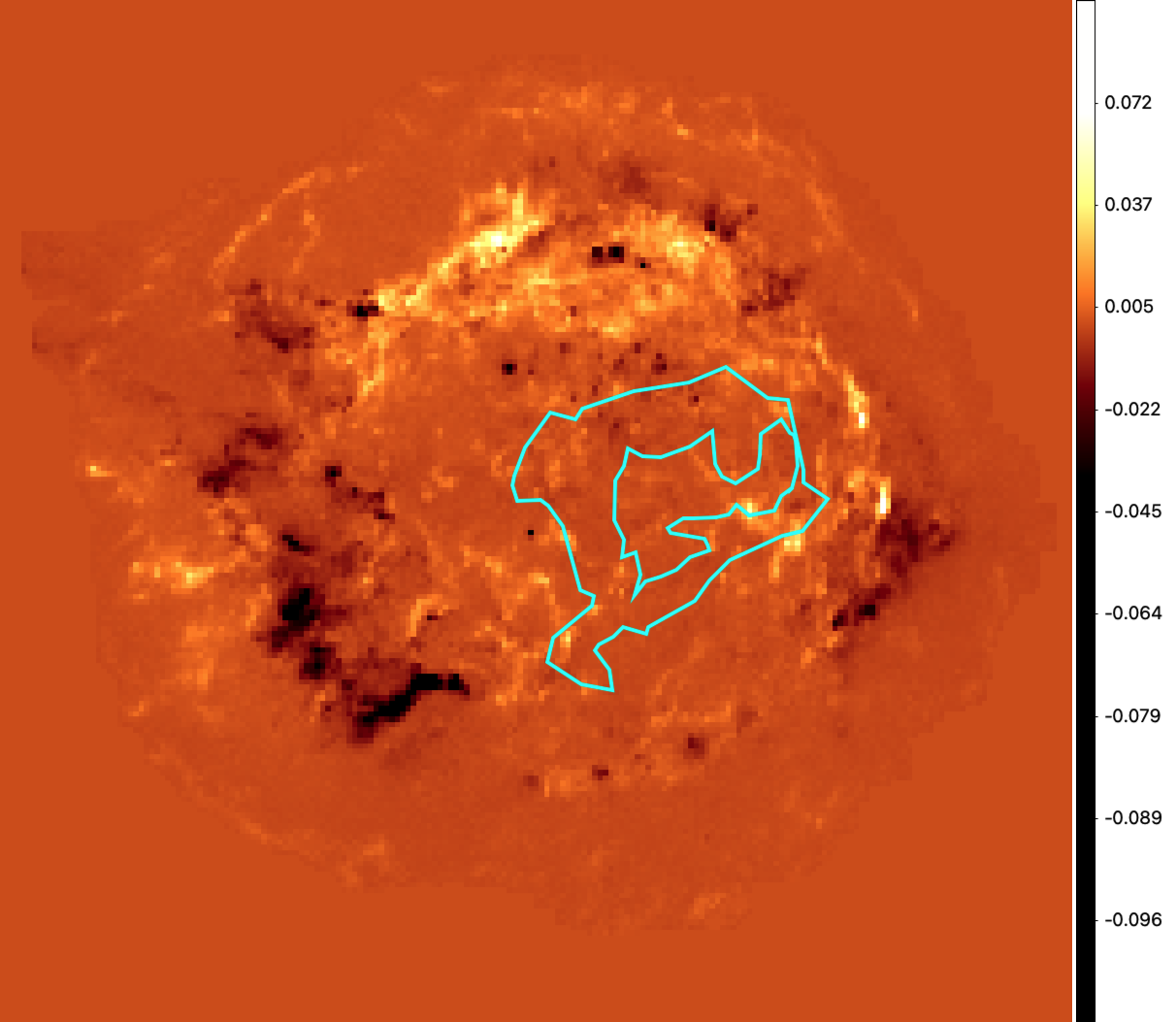}
}
\centerline{
\includegraphics[trim=0 0 0 0,clip=true,width=0.25\textwidth]{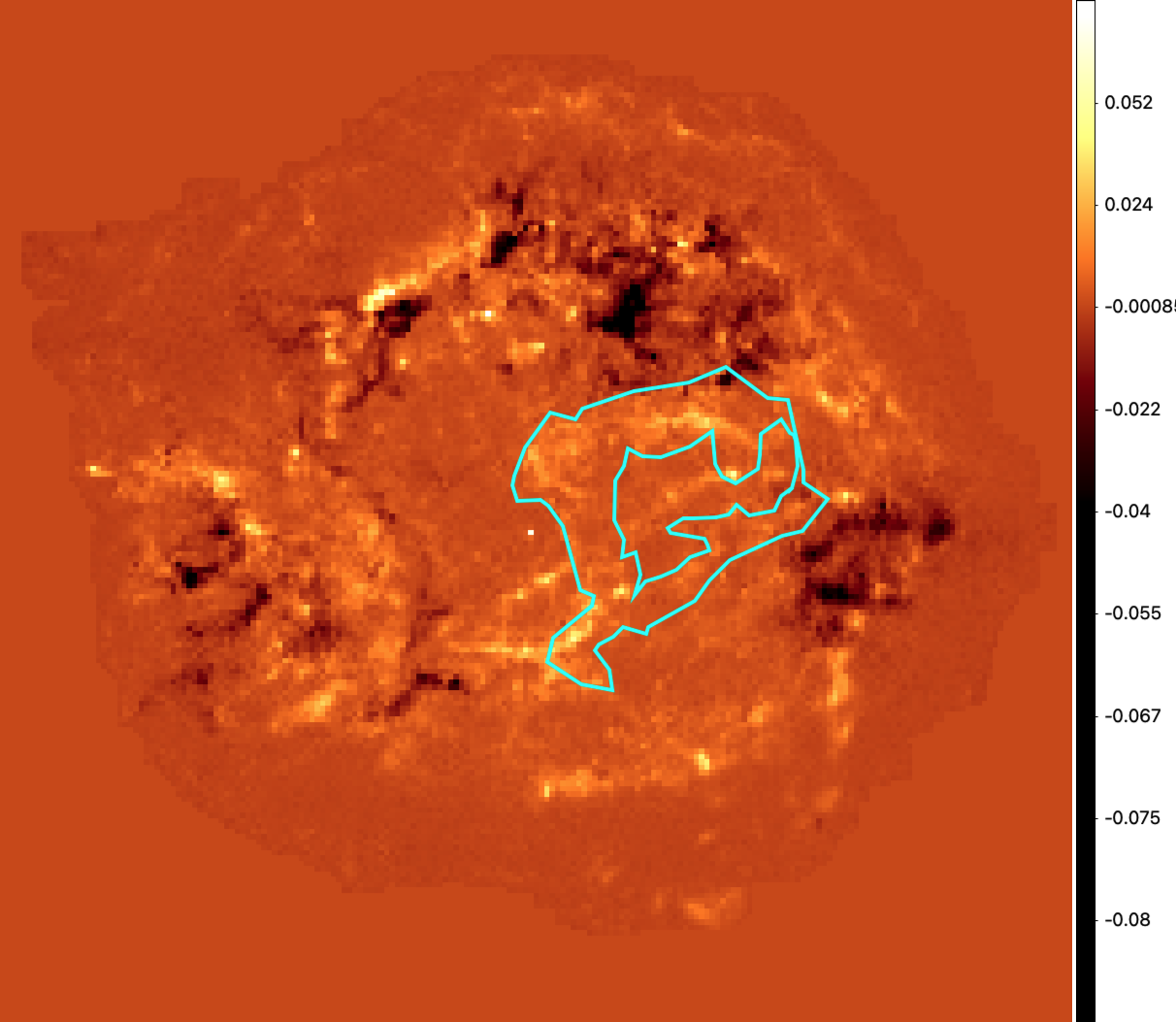}
\includegraphics[trim=0 0 0 0,clip=true,width=0.25\textwidth]{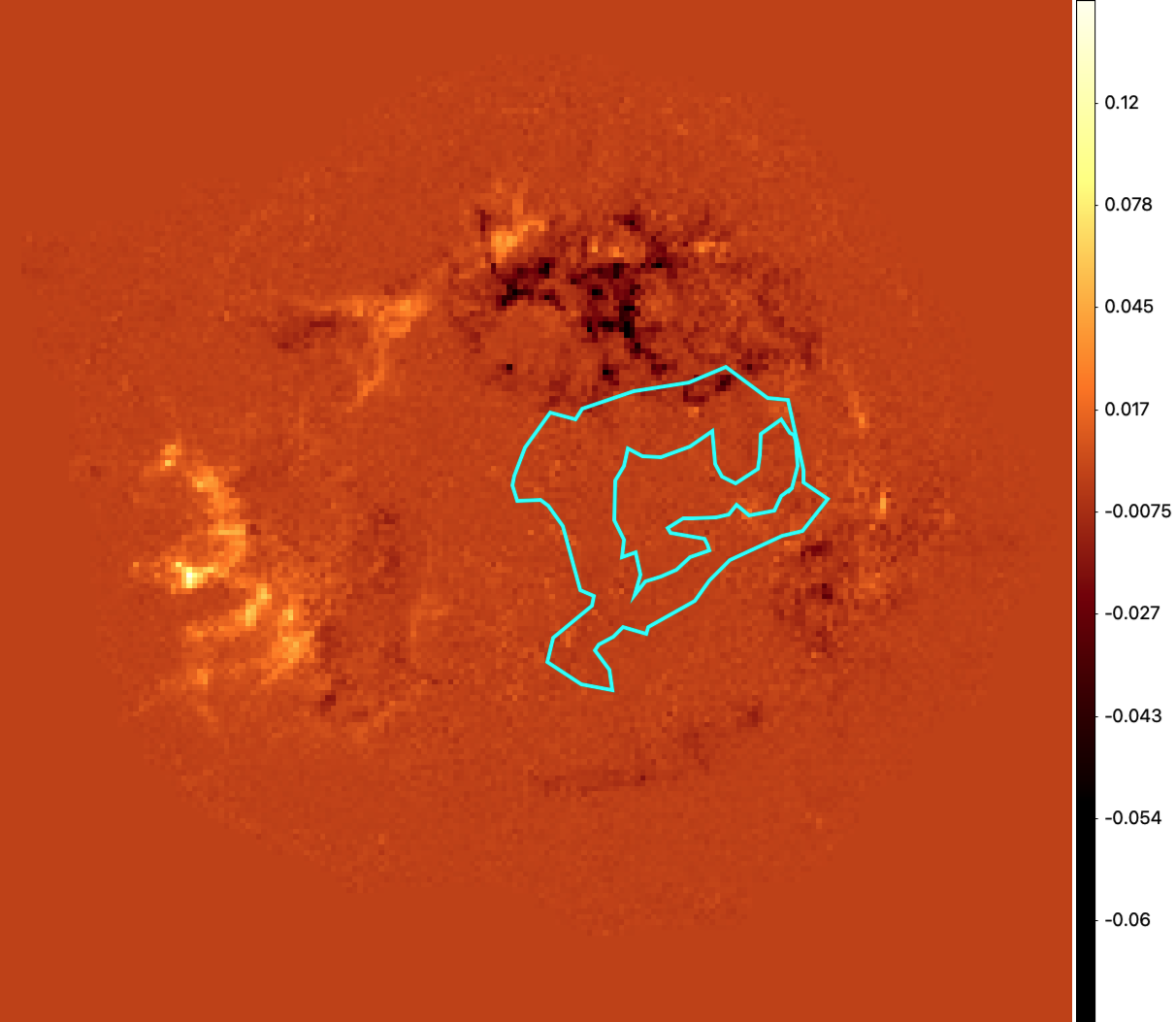}
\includegraphics[trim=0 0 0 0,clip=true,width=0.25\textwidth]{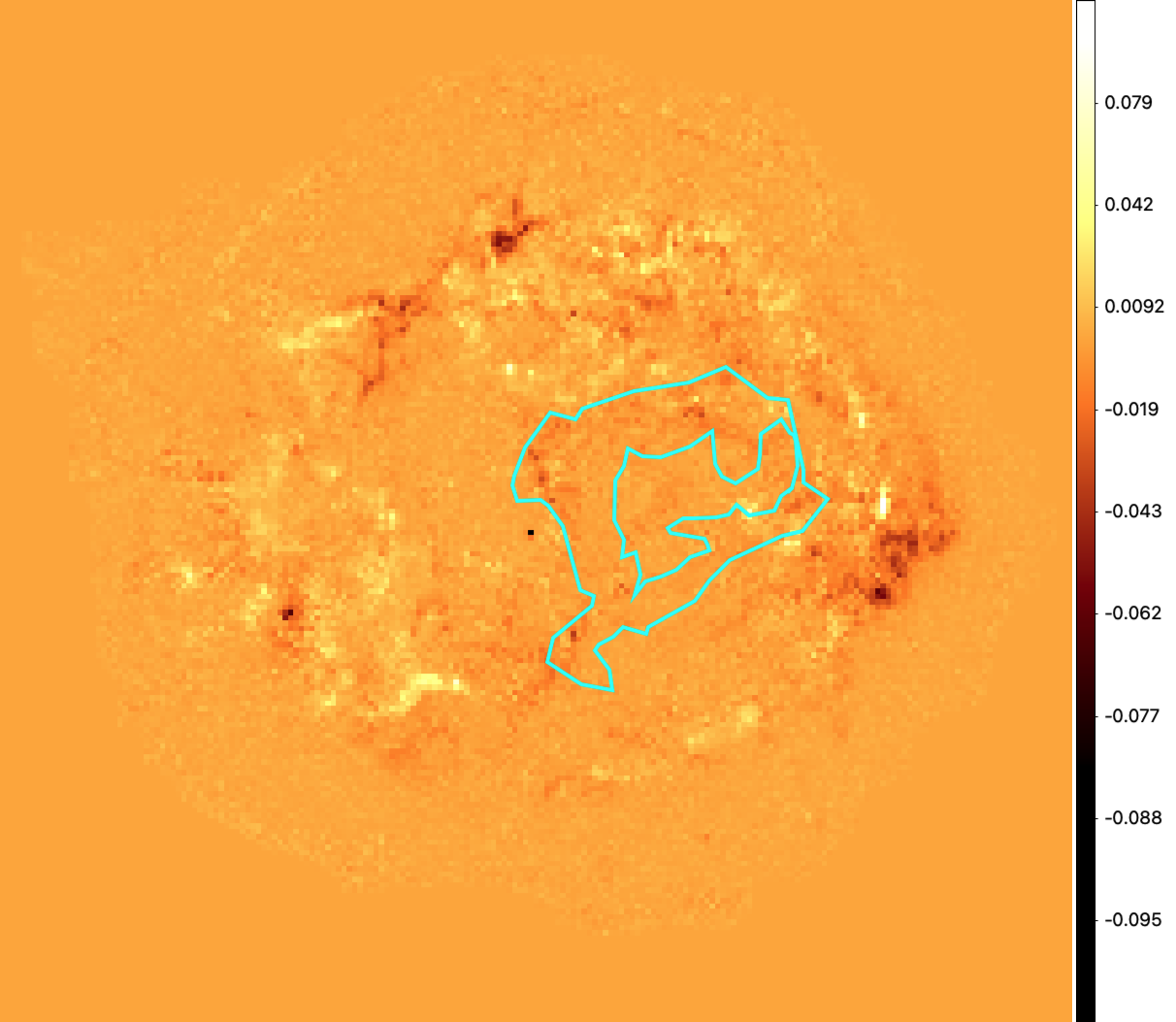}
\includegraphics[trim=0 0 0 0,clip=true,width=0.25\textwidth]{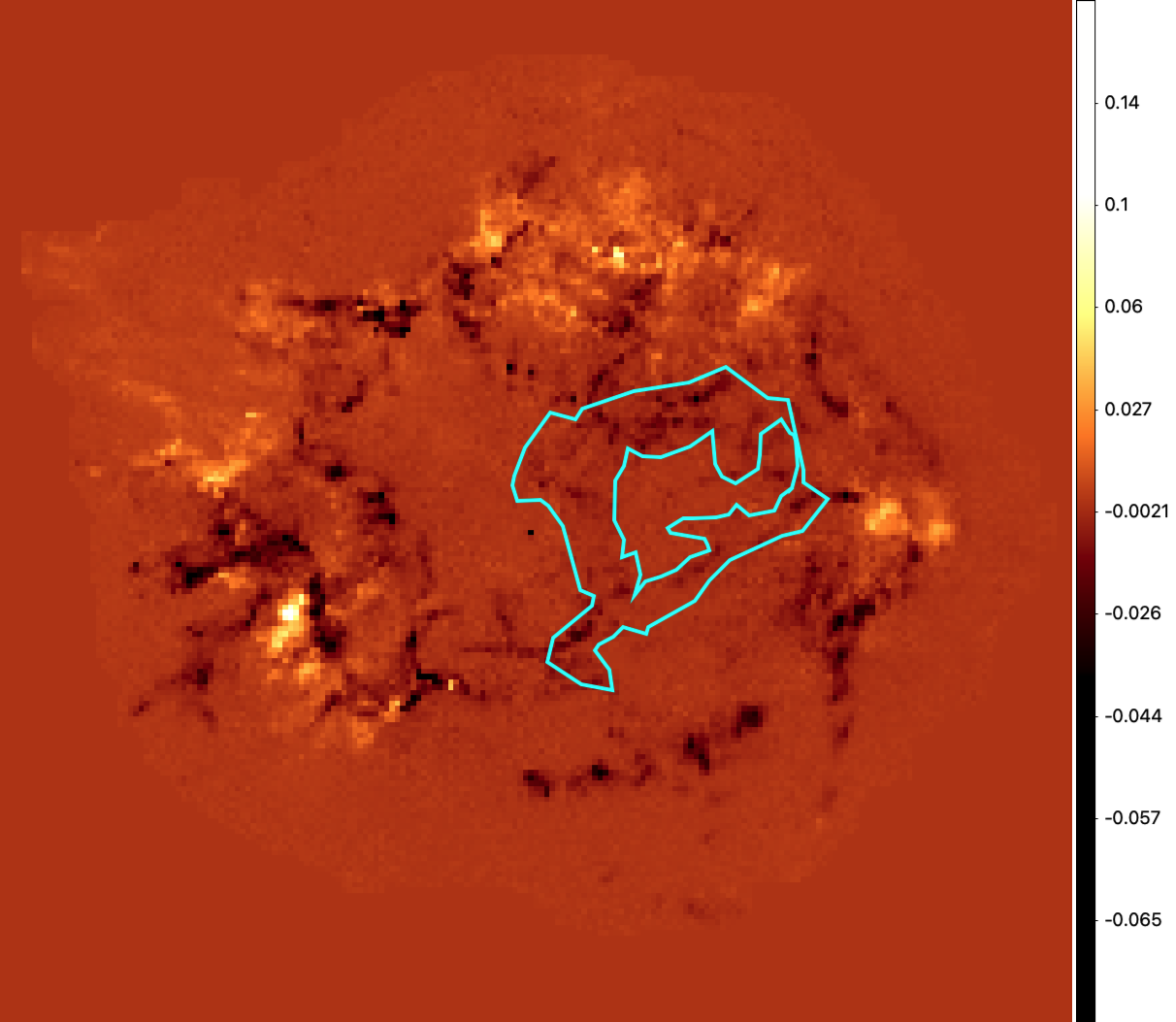}
}
\caption{The PC scores 1 (top left) to 8 (bottom right). 
In particular, in the scores for PC 5, 7, and 8 one recognizes some of the filaments associated with the GM.
\label{fig:pcscores}
}
\end{figure*}

This appendix shows supplementary material that support the main text.
Fig.~\ref{fig:casa_legend} identifies several regions of interest, mentioned in the text.
Fig.~\ref{fig:cornerplots} provides the sampling plots for the spectral fits for the Doppler shift parameters
versus $\tau_{\rm max}$.
Table~\ref{tab:pca} shows the image energy bands used for the PCA.
The first eight PCA score images are shown in Fig.~\ref{fig:pcscores}.

\begin{table}												
\caption{Input energy bands used for principal component analysis. \label{tab:pca}												
}												
\begin{tabular}{rll|rll}		\\\hline\hline\noalign{\smallskip}										
	&	$E_{\rm min}$	&	$E_{\rm max}$	&		&	$E_{\rm min}$	&	$E_{\rm max}$	\\	
	&	(keV)	&	(keV)	&		&	(keV)	&	(keV)	\\ \noalign{\smallskip}\hline\noalign{\smallskip}	
1	&	0.50	&	0.62	&	15	&	2.63	&	2.78	\\	
2	&	0.62	&	0.72	&	16	&	2.78	&	3.00	\\	
3	&	0.72	&	0.87	&	17	&	3.00	&	3.28	\\	
4	&	0.87	&	1.03	&	18	&	3.28	&	3.60	\\	
5	&	1.03	&	1.12	&	19	&	3.60	&	3.75	\\	
6	&	1.12	&	1.23	&	20	&	3.75	&	4.04	\\	
7	&	1.23	&	1.42	&	21	&	4.04	&	4.50	\\	
8	&	1.42	&	1.56	&	22	&	4.50	&	5.00	\\	
9	&	1.56	&	1.69	&	23	&	5.00	&	5.40	\\	
10	&	1.69	&	1.85	&	24	&	5.40	&	5.75	\\	
11	&	1.85	&	1.98	&	25	&	5.75	&	6.25	\\	
12	&	1.98	&	2.13	&	26	&	6.25	&	6.62	\\	
13	&	2.13	&	2.28	&	27	&	6.62	&	6.85	\\	
14	&	2.28	&	2.63	&	28	&	6.85	&	7.35	\\ \noalign{\smallskip}\hline\noalign{\smallskip}	
\end{tabular}												
\end{table}

\begin{figure*}
\centerline{
\includegraphics[trim=300 240 350 300,clip=true,width=0.25\textwidth]{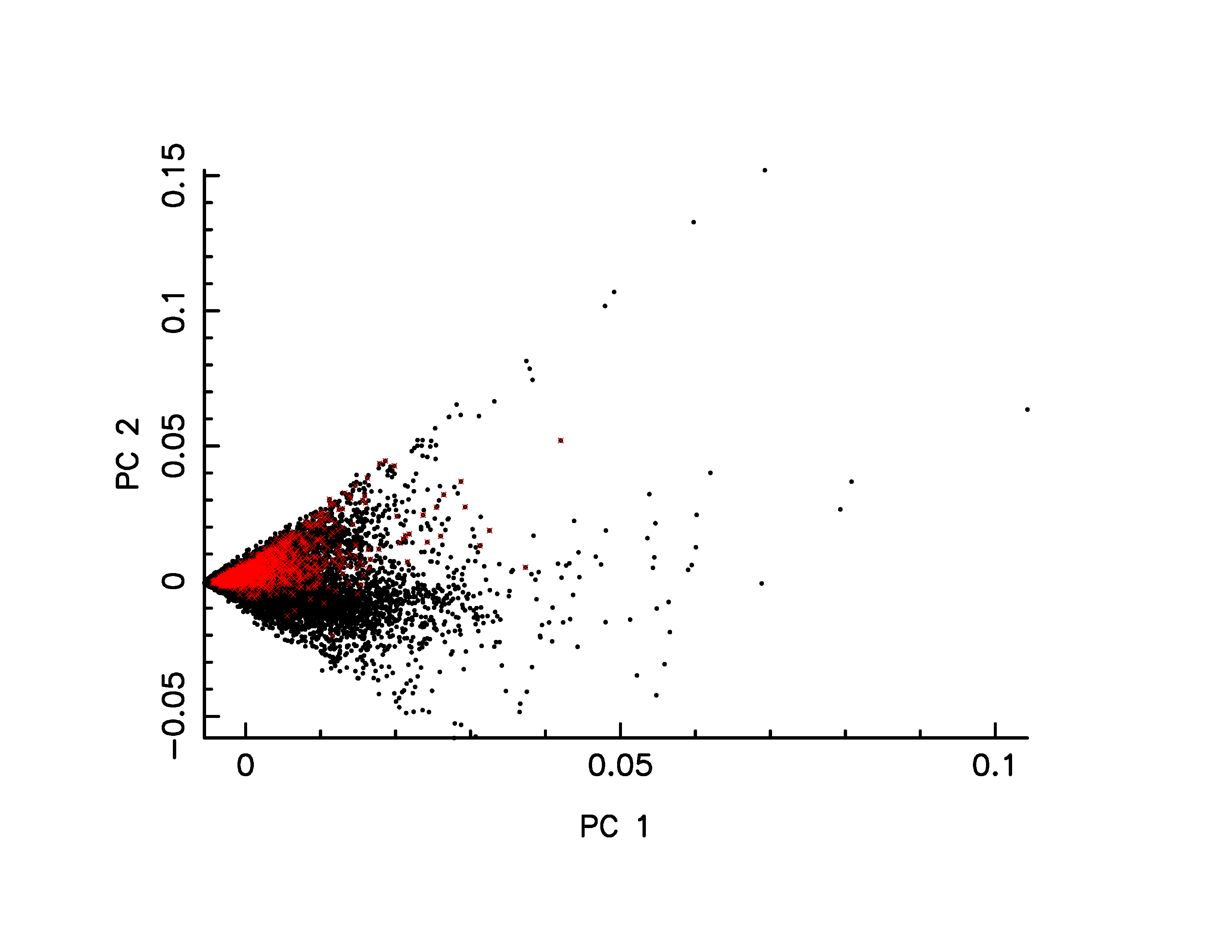}
\includegraphics[trim=300  240 350 300,clip=true,width=0.25\textwidth]{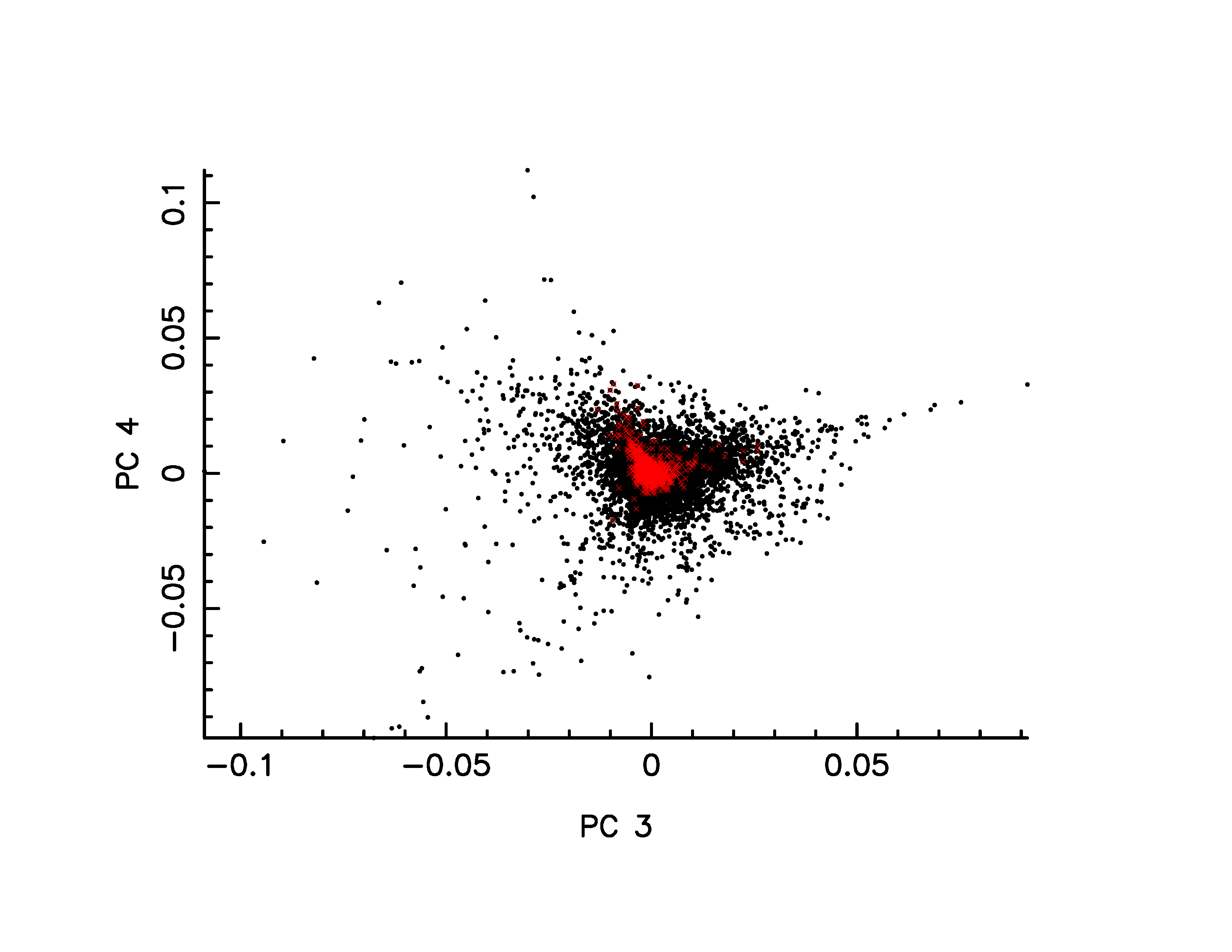}
\includegraphics[trim=300  240 350 300,clip=true,width=0.25\textwidth]{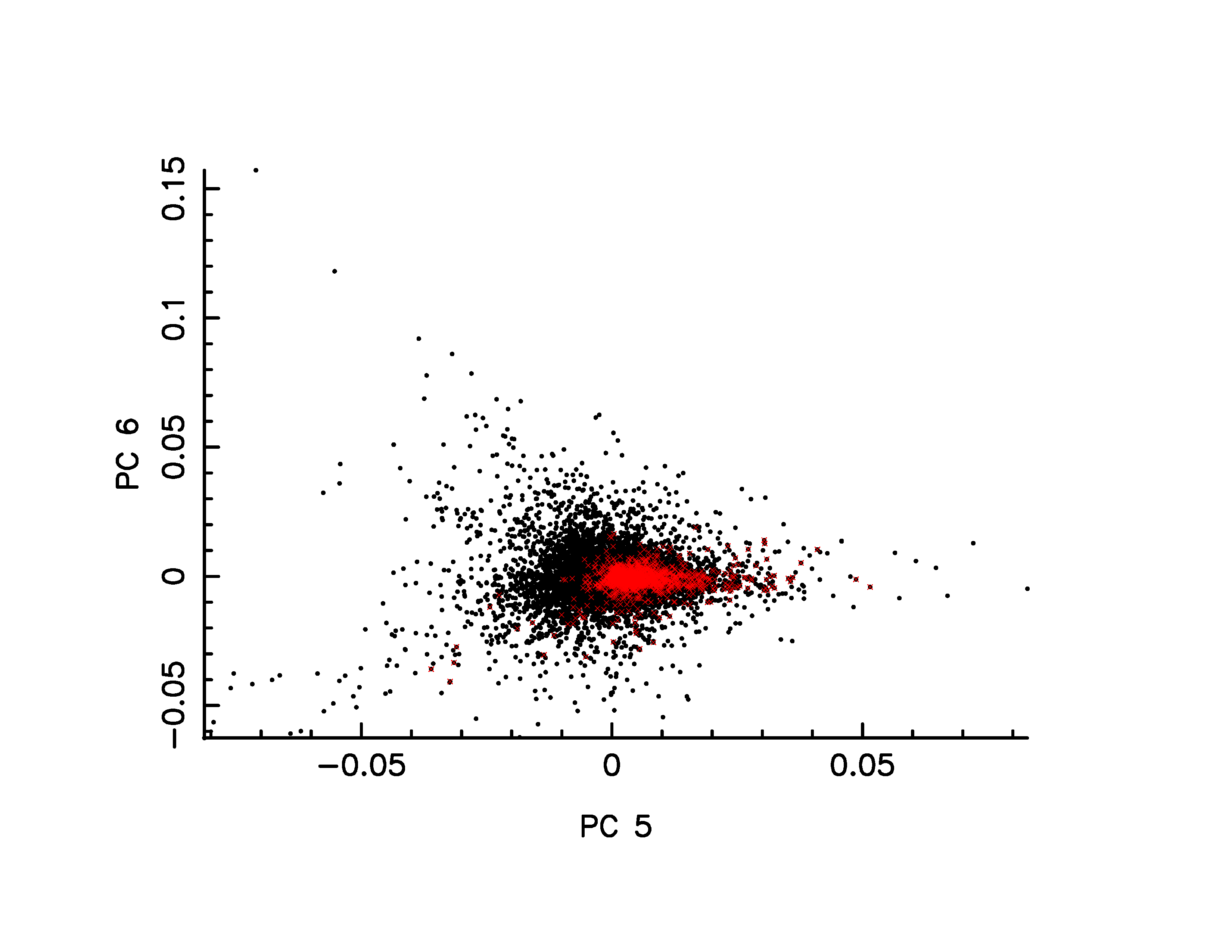}
\includegraphics[trim=300  240 350 300,clip=true,width=0.25\textwidth]{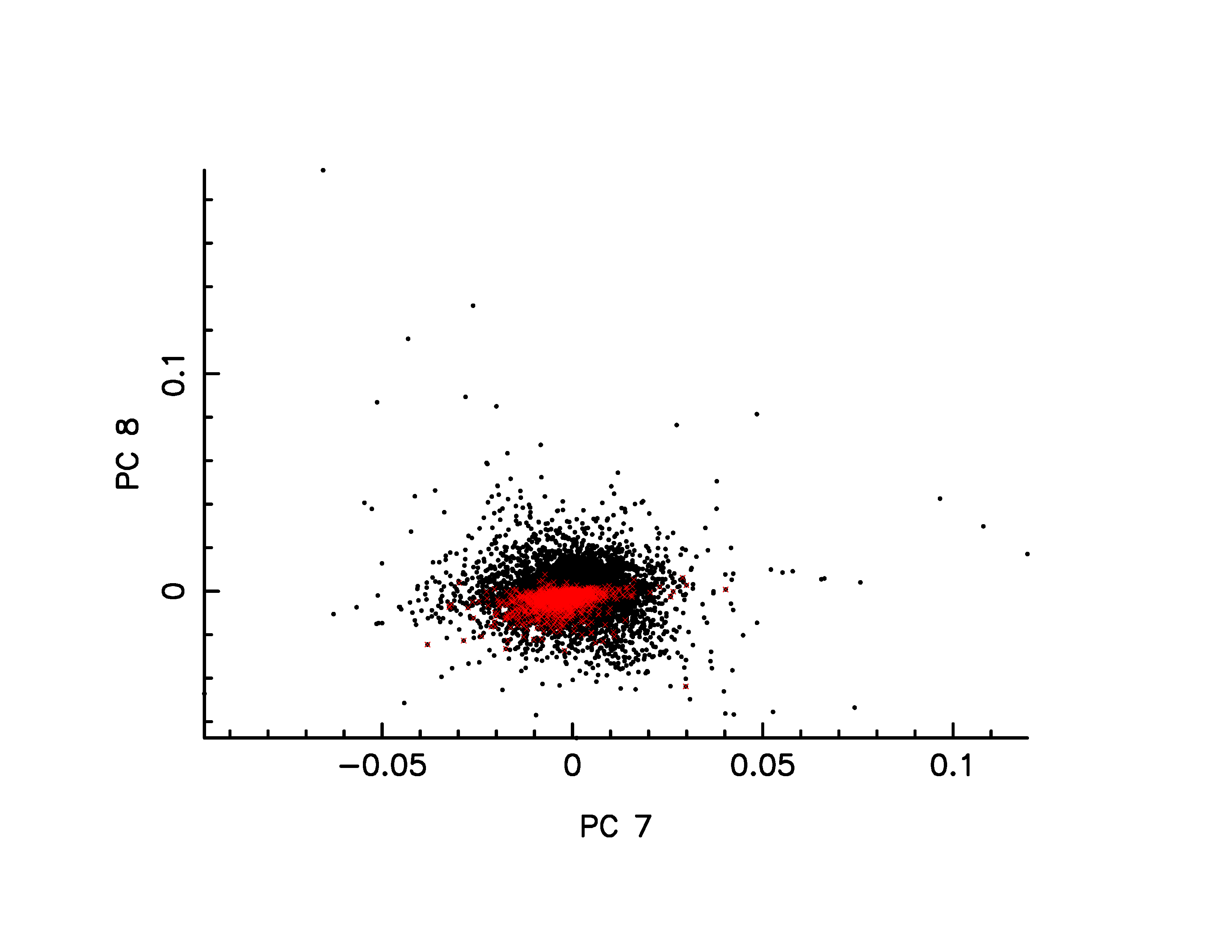}
}
\caption{
Same as top panels of Fig.~\ref{fig:pcscatter}, but now with the full range of PC scores.
\label{fig:pcscatter_full}
}
\end{figure*}

\end{document}